\documentclass[a4paper,11pt]{article}

\usepackage{amsmath,amssymb}
\usepackage[latin1]{inputenc}
\usepackage[authoryear,comma,longnamesfirst,sectionbib]{natbib}
\usepackage[T1]{fontenc}
\usepackage{url}  

\usepackage{lmodern} 
\usepackage{graphicx}
\usepackage{array}
\usepackage{amsfonts}
\usepackage{xcolor}
\usepackage{color}
\usepackage{textcomp}
\usepackage{amssymb} 

\newcommand{\h}{\mathcal{H}}
\newcommand{\N}{\mathbb{N}}
\newcommand{\R}{\mathbb{R}}
\newcommand{\LL}{\mathbb{L}}
\newcommand{\po}{\mathbb{P}}
\newcommand{\E}{\mathbf{E}}
\newcommand{\bias}{\mathsf{bias}}
\newcommand{\var}{\mathsf{var}}
\newcommand{\risk}{\mathsf{risk}}
\newcommand{\hbias}{\widehat{\mathsf{bias}}}
\newcommand{\hvar}{\widehat{\mathsf{var}}}

\newtheorem{theo}{Theorem}
\newtheorem{prop}{Proposition}

\newtheorem{lem}{Lemma}
\newtheorem{df}{Definition}
\newtheorem{remark}{Remark}

\newcommand{\fact}[1]{#1\mathpunct{}!}

\newcommand{\cqfd}{\hfill{$\Box$}}

\renewcommand{\tilde}{\widetilde}
\renewcommand{\hat}{\widehat}

\newcommand{\mathe}{\mathrm{e}} 
\newcommand{\ud}{\mathrm{d}} 
\newcommand\indicator{\mathbf{1}}
\newcommand\ind{\indicator}

\DeclareMathOperator*{\argmin}{arg\,min}

\newcommand\trace{\mathrm{trace}}

\title{A truncation model for estimating Species Richness}
\author{Fran\c{c}ois Koladjo$^{1}$\thanks{Corresponding author. Email address: francois.koladjo@gmail.com}, Mesrob I. Ohannessian$^{2}$, Elisabeth Gassiat$^{3}$}
\date{\vspace{-6ex}}

\begin{document}
\maketitle
\begin{center}
$^{1}$Universit\'e de Parakou, ENSPD BP 55 Tchaourou,   B\'enin   \\
$^{2}$Toyota Technological Institute at Chicago\\
$^{3}$Laboratoire de Math\'ematiques d'Orsay, Univ. Paris-Sud, CNRS, Universit\'e Paris-Saclay, 91405 Orsay, France.  
\end{center}

\begin{abstract}
We propose a truncation model for abundance distribution in the species richness estimation. This model is inherently semiparametric and incorporates an unknown truncation
threshold between rare and abundant counts observations. Using the conditional likelihood, we derive a class of estimators for the parameters in the model by a stepwise
maximisation. The species richness estimator is given by the integer maximising the binomial likelihood when all other parameters in the model are know. Under regularity
conditions, we show that the estimators of the model parameters are asymptotically efficient. We recover the Chao$^{'}$s lower bound estimator of species richeness when the
model is a unicomponent Poisson$^{'}$s model. So, it is an element of our class of estimators. In a simulation study, we show the performances of the proposed method and
compare it to some others.
\end{abstract}

\section{Introduction}

We consider the ``species richness'' problem, also known as the problem of estimating the number of species, which arises when a sample of individuals is taken from a population with $N$ classes or species. The usual data set is a series of observed counts $X^+_1, \dots, X^+_D,$ with $D \leq N$ being the total number of distinct species observed in the sample and $N$ is the parameter to be estimated. Estimating $N$ using such abundance data is an old problem  that has been tackled in several ways, both by parametric models, including Bayesian models (\cite{BUBA08,BABU08}), and by nonparametric models (\cite{WA10}). Due to their flexibility to account for heterogeneity, the nonparametric approaches are those predominantly considered in the last two decades. This setting contains among others the Chao-type estimators developed by Chao and collaborators (see for example~\cite{ChL92,ChB02,ChLo12}), and the likelihood-based nonparametric estimators of which one can cite \cite{NP96,NP98}

Many of these methods, although theoretically founded on a single model, perform the common practice of truncating the data into abundant and rare species. One then assumes that the number of abundant species is adequately represented by the number of distinct such species, whereas the same number leads to an underestimate for the rare species and thus necessitates a correction. Such truncation is generally justified on the basis of avoiding instability. This, however, forces even initially nonparametric models to become effectively parametric, while losing the original hypothesis and the accompanying theoretical guarantees. This motivated us to study this heuristic in a more rigorous light. In particular, we make the following contributions:

\begin{itemize}

\item We give an explicit semiparametric model to represent this truncation practice, where the abundant species are represented by an arbitrary abundance distribution whose support is offset away from the rare range. We partially motivate this as arising from the commonly used Poisson mixtures as being inappropriate for modeling more abundant species.

\item We show that the practice of pure truncation as described above is justified only when the abundant and rare species have abundance distributions whose supports are disjoint. In this case truncation leads to an efficient estimation of the number of species.

\item In general, although pure truncation is not efficient, accounting for the support overlap leads to a hybrid truncation that is a semiparametric procedure which is efficient. We show this by using standard single-parameter families to derive a local minimax bound and a matching (asymptotically) efficient estimator. Coincidentally, we show that this framework recovers several previously suggested estimators as special cases.

\item When the abundance threshold is not known, neither pure truncation nor the hybrid approach can be used directly. For this reason, the proper offset should be obtained from data. We present a model selection approach to resolve this problem. Our experiments show that this approach adapts to the true unknown offset, in the sense that the resulting estimator achieves (almost) the same asymptotic performance as knowing the offset.

\item We illustrate this estimator on both synthetic and real data, showing that our more refined analysis leads to practical improvement.

\end{itemize}

\section{Model and Estimator}

\subsection{Problem Statement}

Assume that $N$ species exist in nature and that each is represented by $X_1,\cdots,X_N$ individuals in a sample. We call $X_i$ the \emph{abundance} of species $i$ in the sample. A classical statistical model of the abundances is to assume that they are independent  random variables identically distributed according to a distribution $f_\nu(x)$ for $x\in {\mathbb N}$ and where $\nu$ is an index within a class of abundance distributions. One of the more common choices of abundance distribution classes are Poisson mixtures indexed by a mixing distribution $\nu$ on $\R_+$:
\begin{equation} \label{eq:poisson-mixture}
f_\nu(x) = \int \frac{\lambda^x\mathe^{-\lambda}}{x!} \ud \nu(\lambda),\quad \textrm{for } x\in {\mathbb N}.
\end{equation}

Of course, we do not get to access non-observed species, i.e. species for which $X_i=0$. If we let $D$ denote the number of distinct observed species, i.e. $D=\sum_{i=1}^N \ind\{X_i>0\}$, and if we re-index and relabel those species as $X^+_1,\cdots,X^+_D$, then it is easy to show that these observed abundances are independent and identically distributed according to the zero-truncated distribution:
\begin{equation}\label{eq:zero-truncation}
f_\nu^+(x) = \frac{f_\nu(x)}{1-f_\nu(0)},\quad \textrm{for } x\in\N_+.
\end{equation}

The central problem of this paper is that of estimating the number of species $N$ with a functional $\hat{N}(X^+_1,\cdots,X^+_D)$ of the abundances of the observed species. In other words, $\hat{N}$ needs to complement the number of observed species $D$ with an estimate of the number of non-observed species.

As outlined in the introduction, a long line of research has addressed this problem. But we focus here in particular on a sequence of influential papers (\cite{ChL92, ChYa93}), the methodology of which continues to be used in more recent papers such as \cite{ChB02} and \cite{WL05}. In theory, these results are within the current framework but, in practice, the estimation is done as follows. The data is divided into rare and abundant components according to an \emph{abundance threshold} $\tau$. Although their estimators are derived and analyzed under the general model, the theoretical estimators are fed with only those abundances such that $X^+_i\leq\tau$, to yield an estimator of the number of rare species $\hat{N}_\mathrm{rare}$. For the abundant species, they use the trivial estimator:
$$
\hat{N}_\mathrm{abundant}=\sum_{i=1}^D \ind\{X^+_i>\tau\}.
$$

The estimate for the total number of species is then simply the sum of both:
$$
\hat{N} = \hat{N}_\mathrm{rare} + \hat{N}_\mathrm{abundant}.
$$

What is the justification behind such truncation? This paper strives to answer this question and to give a more principled model of this common practice, thus leading to a more transparent methodology.

\subsection{Truncation Model}

To motivate the reason behind truncation, note that the justification often given in this line of work (\cite{ChL92, ChYa93, ChB02, WL05}) is that including the abundant species into the estimator may cause instabilities. We can interpret this as the abundance sampling model, and in particular the Poisson mixture model, as not being a good model for abundant species. In this section, we first give some informal insight as to why this may be the case. We then proceed to present an explicit model to handle this rare-abundant dichotomy.

The abundance model can be traced back to a simple sampling model where individuals are drawn independently and identically (with replacement) from a population, where the frequency of species $i$ is $p_i$. If $m$ such individuals are drawn, let $Y_1,\cdots,Y_m$ denote their species. In this model, the abundance of species $i$ has therefore a binomial distribution of parameters $m$ and $p_i$:
\begin{equation} \label{eq:binomial}
X_i=\sum_{j=1}^m \ind\{Y_j=i\} \sim \mathrm{binomial}(m,p_i).
\end{equation}

If the species are not labeled a priori, which corresponds to a random permutation among the $N$ species, then the distribution of a particular abundance becomes a mixture of binomial distributions, with mixture weights at $(m,p_i)_{i=1,\cdots,N}$. Note that these abundances are not independent as in the abundance model, but are rather exchangeable. Notwithstanding this fact, we can see that the abundance model of Equation \eqref{eq:poisson-mixture} effectively replaces this binomial mixture with a Poisson mixture, which cannot be accurate for abundant species.

The source of the instability is due to the fact that a Poisson distribution with a large mean places much more mass near $0$ compared to a corresponding binomial. More precisely, if the model substitutes a binomial mixture with a Poisson mixture, then when an estimator places a mixing mass at a higher abundance, it contributes more to $f_\nu(0)$ than a binomial would. This is then interpreted as evidence of more unseen species than the reality, and thus $N$ is overestimated. This is indeed what is observed with such estimators: with larger values of the truncation $\tau$, the estimate of $N$ tends to increase (see for example the last three columns of Table 2, page 949, and the last two columns of Table 13, page 956, in \cite{WL05}). That said, simply truncating the data is not a theoretically sound approach since the resulting samples no longer follow the hypothesized model. For example Poisson distributions place a positive mass, even if small, beyond any threshold. There is therefore a need to rigorously model rare species, say with mixtures of Poisson distributions, while capturing the possibility that there may be abundant species that have much less influence on our inference about the rare species.

In this paper, we propose the following semiparametric alternative. Let $\tau \in \N_+$ and let $\mathcal{F}_{\tau}$ be the family of discrete distributions supported on $\{\tau+1,\tau+2,...\}$. We assume that abundances follow a distribution $f$ that belongs to a model $\mathcal{P}$, as follows. For $\alpha>0$ define:
\begin{equation}\label{Eq:Model}
\mathcal{P} = \left\{f_{(q,\theta,F)}(x) = qR_{\theta}(x) + (1-q)F(x)  \right\}
\end{equation}
with $q \in [\alpha,1)$, where $R_{\theta}$ is a parametric model that represents mostly rare species (e.g. we may think of a finite mixture of Poisson distributions, and more generally we ask for $\theta \in \varTheta$, where $\varTheta$ is an appropriate subset of $\R^k$ for some $k\in \N_+$), and where $F \in \mathcal{F}_{\tau}$ is a nonparametric component that represents abundant species.


Using Equation \eqref{eq:zero-truncation} and the fact that the nonparametric component vanishes at $x=0$, this model induces the zero-truncated version as follows:
\begin{equation}\label{Eq:Model_t}
  \mathcal{P}^{+} = \left\{ f_{(q,\theta,F)}^{+}(x) = \frac{f_{(q,\theta,F)}(x)}{1-qR_{\theta}(0)},
  f_{(q,\theta,F)} \in \mathcal{P}\right\}.
\end{equation}
We leave the choice of $R_\theta$ open, except for certain identifiability and smoothness assumptions that we later spell out in detail. Thus $R_\theta$ is not necessarily a Poisson mixture.
The choice of a parametric model for $R_\theta$ is justified by the fact that even originally nonparametric models are effectively reduced to parametric classes under the constraint of identifiability from a small (truncated) support.

It is now clear that our model in Equation \eqref{Eq:Model_t} makes explicit the notion that rare and abundant species may coexist. This allows us to bypass heuristics and suggest estimators with provable performance guarantees. In particular, we may harness the basic theory of semiparametric models to establish the efficiency of likelihood-based estimators, and suggest potential model selection mechanism for the choice of $\tau$. Furthermore, as we make no further assumptions beyond adopting a parametric form for the rare component and dislocating the support of the abundant species away from zero, we have a model that can go beyond a simple justification of truncation. For example, one may think of $F$ as a nonparametric corruption to the data, rather than a legitimate measurement of abundant species, and our analysis and methodology still goes through unaffected.

\subsection{Estimator of the Number of Species}

The estimator that we propose for $N$ falls under the category of maximum likelihood (MLE)-type M-estimators. In this section we derive and define the estimator, and in the next section we study some of its asymptotic properties.

Let $n_{x} = \sum_{i=1}^{D} \ind\{X_{i}^{+}=x\},$ $x \geq 1$, be the empirical counts of the abundances, which are sufficient statistics for computing likelihoods. The combined likelihood of $N$ and the rest of the model parameters given the samples can then be written as follows:
\begin{equation} \label{Eq:Vraiss}
  L(N, f|(n_{x})_{x\geq 1}) = \dfrac{N!}{(N-D)!\prod_{x}n_{x}!}f(0)^{N-D}
  \prod_{x\geq1}f(x)^{n_{x}},
\end{equation}

It is interesting to note that this likelihood may be decomposed into the product of two likelihoods. The first is the likelihood of $N$ given the rest of the model parameters and the number of distinct samples $D$. This has a binomial form and we denote it by $L_b$. The second is the likelihood of the rest of the model parameters, given the samples, and we denote it by $L^{+}$. After substituting $f(x)$ by its expression in $\mathcal{P}$, in particular letting $f(0)=qR_\theta(0)$, and noting that $D=\sum_x n_x$, we can write these likelihoods respectively as follows:
\begin{equation} \label{Eq:Lb}
  L_{b}(N|D,q,\theta) = \frac{N!}{D!(N-D)!}
  \left[qR_{\theta}(0)\right]^{N-D}\left[1-qR_{\theta}(0)\right]^{D},
\end{equation}
and
\begin{equation} \label{Eq:L+}
  L^{+}(q,\theta,F|(n_{x})_{x\geq 1}) = \frac{D!}{\prod_{x\geq1}n_{x}!}
  \prod_{x\geq1}{\underbrace{\left[\frac{qR_{\theta}(x) + (1-q)F(x)}{1-qR_{\theta}(0)}\right]}_{f^+(x)}}^{n_{x}},
\end{equation}
suggesting two methods to undertake the maximum likelihood estimation from $L$. Note that some of the earliest works to suggest such a decomposition were~\cite{SAN71,SAN77} (see also~\cite{ML03,ML07} for a more recent treatment).

The first method is to maximize directly the likelihood $L$ over all of $(N,q,\theta,F).$ The estimator of $N$ obtained from this method is typically called the \emph{unconditional} maximum likelihood estimator. For example, some nonparametric models with unconditional estimation methods are proposed in~\cite{NP96,BD05}. The second method to obtain a maximum likelihood estimator of $N$ is to first maximize the likelihood $L^{+}$ from the zero-truncated model $\mathcal{P}^{+}$ to derive the estimators of $q,$ $\theta$ and $F,$ and then to maximize the binomial likelihood $L_b$ in the parameter $N$ given that $q$ and $\theta$ are known. This method is known as the \emph{conditional} maximum likelihood method for estimating $N$.

We consider here only the conditional maximum likelihood method. Before we proceed with the estimation of $\theta$, $q$, and $F$, note that maximizing the binomial likelihood $L_b$ in Equation \eqref{Eq:Lb} gives us the form of our estimator:
\begin{equation} \label{Eq_mleN}
 \hat{N}(q,\theta) = \dfrac{D}{1-qR_{\theta}(0)}.
\end{equation}

The final expression for the estimator therefore consists of estimating $\theta$ by $\hat\theta$ and $q$ by $\hat q$, in a manner that we shortly outline, and then substituting in Equation \eqref{Eq_mleN} to obtain $\hat{N} = \hat{N}(\hat{q},\hat{\theta})$. Of course, $N$ is an integer parameter, and we could then take the integer part of the resulting estimate. That said, in what follows we allow ourselves to accept non-integer estimates.

We now proceed to estimate the parameters. We observe first that since $F$ plays no role in the expression for $\hat{N}$, we can treat it as a nuisance parameter. The next observation is that to maximize $L^{+}$, we can successively fix some parameters while we maximize over others. Because $F$ is mostly a nuisance parameter, we maximize the likelihood $L^{+}$ when $q$ and $\theta$ are fixed without further constraining $F$ to be a proper distribution. This approach gives us, at each support point $x$, the following pseudo-estimator, as a function of $\theta$ and $q$:
\begin{equation} \label{mleF}
\hat{F}(q,\theta)(x) = \frac{[1-q\sum_{k=0}^{\tau}R_{\theta}(k)]}{(1-q)(D-D_{\tau})}n_{x} -
\frac{q}{1-q}R_{\theta}(x),
\end{equation}
where $D_{\tau}=\sum_{x=1}^\tau n_x$ denotes the number of species with abundance no greater than $\tau$.

The reason we call $\hat{F}(q,\theta)$ a pseudo-estimator is that it may put negative mass at some of its support points as it is not constrained to be nonnegative. This occurs for example at the non-observed support points of $F$, that is for a support point $x$ such that $n_{x}=0$. Despite this fact, the estimators for $\theta$ and $q$ that follow from this choice of $\hat{F}$ are not sensitive to its impropriety.


Replacing $F$ by its pseudo-estimator in the expression for $L^{+}$ leads to an objective function for $q$ and $\theta$ which may now be maximized in $q$. This leads to an MLE-type estimator of $q$, still as a function of $\theta$:
\begin{equation}\label{eq:q-hat}
  \hat{q}(\theta) = \frac{1}{R_{\theta}(0)+ \frac{D}{D_{\tau}} \sum_{k=1}^{\tau} R_{\theta}(k)}.
\end{equation}
Note that $q$ is always non-negative. However, for particular values of $\theta$, $D$, and $D_\tau$, it could be larger than $1$. If this occurs in practice, we simply constrain it to $1$ to obtain a valid probability. The consistency result in the next section shows that this is not a concern, asymptotically.

The last step is to find a proper estimator of $\theta$. Consider the following simplifying notation. For a fixed $\tau$, let $S_{\theta}^{\tau}$ denote the truncated version of the density $R_{\theta}$ defined as
\begin{equation}\label{GenTruncated}
  S_{\theta}^{\tau}(x) = \frac{R_{\theta}(x)}{\sum_{k=1}^{\tau}R_{\theta}(k)} \text{ for } 1 \leq x \leq \tau.
\end{equation}
By replacing $F$ and $q$ by their estimators in the conditional likelihood $L^{+}$, we can show that we obtain (up to factors that do not depend on $\theta$) the following truncated likelihood:
\begin{equation}\label{TruncVrais}
 \prod_{x=1}^{\tau} \left\{ S_{\theta}^{\tau}(x)\right\}^{n_{x}}.
\end{equation}

The estimator $\hat \theta$ is then simply a maximizer of Equation \eqref{TruncVrais}. We can thus see that $\hat\theta$ is an MLE of the truncated density $S_{\theta}^{\tau}$, based on the first $\tau$ abundance counts. This completes our estimator construction. Indeed, to estimate $N$, we first compute $\hat\theta$ directly from the samples by maximizing Equation \eqref{TruncVrais}, we then calculate $\hat q(\hat\theta)$ using Equation \eqref{eq:q-hat}, and lastly we substitute both to obtain $\hat N(\hat q(\hat\theta),\hat\theta)$ using Equation \eqref{Eq_mleN}.

We conclude by noting that all the derivations we performed were based on the premise that a value of $\tau$ was given. As $\hat{q}$ and $\hat{\theta}$ depend on $\tau$, in what follows either we make this explicit by writing $\hat{q}_{\tau}$ and $\hat{\theta}_{\tau}$ respectively or keep it implicit when the notation gets encumbered. Similarly we write $\hat{N}_\tau$. We also sometimes use the notation $\hat{q}(\hat{\theta})$ instead of $\hat{q}$ to make it explicit that the estimator of $q$ depends on $\hat{\theta}$.

\subsection{Relationship to Other Estimators}

Despite the fact that $\theta$ is estimated by truncating the model to the abundance values between $1$ and $\tau$, our estimator differs from the traditional truncation with conditional MLE-type estimators often described in the literature, as overviewed in the introduction. To be precise, assume the same parametric rare-species model is used for $R_\theta$, and recall that in these classical estimators the data is truncated and the conditional MLE is solved using the zero-truncated version of $R_\theta$ to obtain $\hat{\theta}$, and then the rare-species count is estimated by:
$$
\hat N_\mathrm{rare} =  \frac{D_{\tau}}{1-R_{\hat{\theta}}(0)}.
$$

The abundant species are then assumed to be represented exactly by what is seen:
$$
\hat N_\mathrm{abundant} = D-D_\tau.
$$

The combined estimator is therefore:
\begin{eqnarray*}
\hat N_\mathrm{classical}
  &=& \hat N_\mathrm{rare} + \hat N_\mathrm{abundant} \\
  &=& \frac{D_{\tau}}{1-R_{\hat{\theta}}(0)} + (D-D_\tau).
\end{eqnarray*}

The following proposition identifies the condition under which our estimator is equivalent to this classical estimator.

\begin{prop} \label{Equiv_MLE}
 If $R_{\theta}$ is supported on $\{0,\ldots,\tau\},$ then the two estimators $\hat{N}_{\tau}$ and $\hat{N}_\mathrm{classical} $ are equivalent.
\end{prop}

Proposition~\ref{Equiv_MLE} means that if the parametric part $R_{\theta}$ and the nuisance parameter in the model $ \mathcal{P}^{+}$ are supported on disjoint sets, then one can split the data set into rare-species data $(X_i\leq \tau)$ and abundant-species data $(X_i > \tau)$. In this context, inference on rare species is not affected by the estimation of the nuisance parameter $F$ and thus throwing away high-abundance data is justified. On the other hand when $R_\theta$ does extend over all integers, then one should not ignore any part of the data, and instead one should perform a hybrid truncation, as suggested by $N_\tau$ in order to obtain efficient estimators.

Thus far we have considered the general context for any eligible distribution $R_{\theta},$ some particular cases enable us to make simple and concrete connections between $\hat{N}_{\tau}$ and other popular estimators that come close to falling within our framework. In particular, Chao, in \cite{Ch84}, suggests the following popular estimator
$$
\hat{N}_\mathrm{Chao} = D + n_{1}^{2}/2n_{2}.
$$
The following proposition shows that our estimator $\hat{N}_{\tau}$, for $\tau = 2$ and $R_{\theta}$ corresponding to a pure Poisson distribution, is equivalent to Chao's $\hat{N}_\mathrm{Chao}$. As such, we can interpret our estimator as a generalization of Chao's, where $\tau$ is no longer restricted to $2$ and where $R_\theta$ may be more general than a pure Poisson distribution.

\begin{prop} \label{Equiv_MLE_Zelt}
  Assume that $ \tau = 2$ and $R_{\theta}(x) = \theta^{x}e^{-\theta}/\fact{x}$ for all $x \geq 0.$ Then $ \hat{N}_{\tau} = \hat{N}_\mathrm{Chao}.$
%
%
%
 \end{prop}

It is worth noting that Zelterman in \cite{Ze88} explicitly considers the pure Poisson model with access only to the first two counts $n_{1}\neq 0 $ and $n_{2}$, and suggested $\hat{\theta}_\mathrm{Zelterman}= 2n_{2}/n_{1}$ as an estimator for $\theta$, showing certain robustness properties under heterogeneity in the true model (see \cite{Ze88} for more details). It is indeed straightforward to verify that for $\tau=2$ and a pure Poisson model for $R_\theta$, we have that our estimator maximizing the truncated likelihood of Equation~\eqref{TruncVrais} corresponds to that of Zelterman: $\hat{\theta}_\tau=\hat{\theta}_\mathrm{Zelterman}$.

An effective proof of Proposition \ref{Equiv_MLE_Zelt} is given in the work of B\"{o}hning et al in~\cite{BVALVA13} within an alternative framework: using the conditional expectation of $f_{0}$. In the Appendix, we propose a simpler proof that is more in line with our framework, by plugging-in $\hat{\theta}_\mathrm{Zelterman}$, which as noted is the correct conditional ML estimate, into our expression for $\hat{N}_{\tau}.$

We end by noting that if the abundant species are not taken into account, we would have the estimator of $N$ given by $\hat{N}_\mathrm{Zelterman}= D/[1-\exp(-\hat{\theta}_\mathrm{Zelterman})]$. This estimator is in the spirit of pure-truncation, and would clearly deviate from $\hat N_\tau$ (the denominator has no $q$ factor). Since we establish the latter to be consistent within our model, then it follows that the former is not (see also~\cite{BH09} for a more quantitative comparison between Chao's and Zelterman's estimators of $N$.)

\subsection{Choice of $\tau$ via Model Selection} \label{sec:model-selection}

 To end the discussion of our estimator of $N$, we stress once again that $\hat N_\tau$ depends on the integer truncation parameter $\tau$, which delimits the zone of influence of the abundant species through the support of the nuisance parameter $F$. When $\tau$ is not known, we need a procedure to estimate this parameter. This is effectively a model selection problem, which we now address using the Goldenshluger-Lepski (G-L) method as inspiration. The G-L method was introduced in~\cite{goldenshluger2011} in the context of bandwidth selection for kernel density estimation. In the current paper we use it heuristically, without formal proofs. Experimental evidence, however, suggests that the method is very effective.

The principle of the method is as follows. As our estimator is of the form $\hat N=D/(1-\hat P(0))$, we focus on the problem of estimating $P(0)=q R_\theta(0)$. Let us drop the $(0)$ argument from the notation, to make the exposition clearer. Assume that we have a known upper bound $\tau_{\max}$ on the largest value $\tau$ could take, and let $\tau_{\min}$ be the least $\tau$ that enables the necessary identifiability assumptions. If we relax the requirement that $F$ is positive on its support, we have successively smaller nested models as $\tau$ varies from $\tau_{\min}$ to $\tau_{\max}$. Each of these models has a corresponding version of our estimator, that we denote by $\hat P_\tau$. The (squared) \emph{bias} of each model is $\bias_\tau=(\E[\hat P_\tau] - P)^2$. The variance of each model is $\var_\tau=\E[(\hat P_\tau-\E[\hat P_\tau])^2]$. The mean squared error \emph{risk} decomposes as usual into the sum of bias and variance,  $\risk_\tau = \E[(\hat P_\tau - P)^2]=\bias_\tau+\var_\tau$. Now observe the following:
\begin{itemize}
 \item For $\tau\leq \tau_0$, the consistency result of Theorem \ref{Consistency} tells us we are asymptotically unbiased.
 \item For $\tau=\tau_0$ Theorem \ref{TheoEfficiency} shows that we are efficient and therefore asymptotically we have the least variance.
 \item For $\tau<\tau_0$, the estimator becomes inefficient and the variance may be higher. Intuitively, this is because less of the data is used to estimate $\theta$ when the truncation is stricter.
 \item For $\tau>\tau_0$, Theorem \ref{Consistency} tells us that we may have a non-vanishing bias. However, the variance itself may be lower simply because more $F$-corrupted data is used to converge to an incorrect value of $\theta$.
\end{itemize}

The inevitable bias-variance tradeoff thus manifests itself in this framework, and the best compromise in terms of risk will be achieved at the correct model class $\tau_0$. If accurate proxies $\hbias_\tau$ and $\hvar_\tau$ are available, then we may empirically select a model $\hat\tau$ near $\tau_0$, by minimizing
\begin{equation} \label{tauh}
    \hat\tau = \argmin_\tau \left(\hbias_\tau+\hvar_\tau\right).
\end{equation}

The bootstrap method is one effective way for estimating $\var_\tau$. In its simplest version, bootstrap consists of resampling $D$ points from the data and computing an estimator $\tilde P_\tau$ from the resampled data. Then this is repeated a number of times, say $j=1,\cdots,M$, and the variance is estimated as:
\begin{equation} \label{eq:bootvar}
    \hvar_\tau = \tfrac{1}{M} \sum_{j=1}^M (\tilde P_{\tau,j}-\hat P_\tau)^2.
\end{equation}

While the resampling process of the bootstrap is good at quantifying the \emph{relative} (to $\hat P_\tau$) variability of the resampled estimators, it offers no \emph{absolute} reference point, crucial for estimating the bias. Luckily, as we have argued, the larger model classes have small bias and can themselves be used as a reference point. The Goldenshluger-Lepski method suggests the following method to obtain a bias proxy:
\begin{equation} \label{eq:bias}
    \hbias_\tau = \max_{\tau' \leq \tau} \left[ (\hat P_{\tau'}-\hat P_\tau)^2 - \hvar_{\tau'} \right]_+,
\end{equation}
where $[\cdot]_+$ stand for the non-negative part. The justification and behavior for this bias proxy needs to be rigorously established, as is done for kernel width selection in \cite{goldenshluger2011}. For our heuristic use, we provide simply the intuition behind it. This formula can be interpreted by noticing that the maximum of $(\E[\hat P_{\tau'}]-\E[\hat P_\tau])^2$ over $\tau'\leq \tau$ is indeed approximately the bias since, as we described, the smaller models are (asymptotically) unbiased. But because we only have access to $(\hat P_{\tau'}-\hat P_\tau)^2$ instead of $(\E[\hat P_{\tau'}]-\E[\hat P_\tau])^2$, and since the smaller models have higher variance, we place a conservative confidence bound on the $\tau'$ end using $\hvar_{\tau'}$ in order not to overestimate the bias.

Equations \eqref{tauh} (the selection of $\hat\tau$), \eqref{eq:bootvar} (the bootstrap variance proxy), and \eqref{eq:bias} (the bias proxy) completely specify a heuristic model selection procedure for estimating the integer truncation parameter $\tau$.

\section{Analysis of the Estimator} \label{sec:analysis}

\subsection{The semiparametric framework}

We now analyze the convergence and optimality of our estimator in the context of efficient estimation, when the model contains nuisance parameters. We do so particularly in order to handle the nonparametric component $F$ within our semiparametric model. In the absence of such nuisance parameters, efficiency may be defined in terms of attaining the Cram\'er-Rao bound. In regular parametric models, the Cramer-Rao bound is the variance of the score function, itself  (often) defined as the derivative of the log-likelihood, and efficient estimators are at first order empirical means of the score function.
The nuisance parameters, however, can lead to unavoidable  loss in the accuracy of any estimator. The notion of efficiency can then be extended by assessing new lower bounds to the variance of the parameters of interest. We provide the details in Section \ref{sec:effi} below, and describe here what is useful to state our results. 

One can define a set $ \mathcal{\dot{P}}_{F}^{+}$ of score functions relatively to the nonparametric part of the model, built using one dimensional submodels (see Section \ref{sec:effi} for details). Then, if $\dot{\ell}_{(q,\theta)}$ is the usual score function (given by the partial derivative and gradient with respect to $q$ and $\theta$ respectively of the log-likelihood in the full model), the \emph{efficient score function} related to $(q,\theta)$ 
is then defined component-wise as $\tilde{\ell}_{(q,\theta)}=\dot{\ell}_{(q,\theta)} - \varPi_{F} \dot{\ell}_{(q,\theta)},$ where $\varPi_{F}$ is the orthogonal projection onto the closure of the linear space spanned by $\mathcal{\dot{P}}_{F}^{+}.$ The efficient score functions play the same role for efficient estimators (if they exist) as the ordinary score functions for the maximum likelihood estimators in a parametric model with no nuisance parameter. Namely, they lead to the best asymptotic variance for any estimator. The corresponding efficient Fisher information $\tilde{I}_{(q,\theta)}$ is a matrix whose components are the variances and covariances of the various components of the vector of efficient score functions.

As such, this leads to what we shall give as formal definition of the properties of consistency and efficiency:
\begin{df}
As $N\to\infty$, an estimator sequence $T_D=(\hat q
, \hat\theta%
)$ is:
\begin{itemize}
 \item \emph{Consistent}, if $T_D\to(q,\theta)$ in probability.
 \item \emph{Efficient} (asymptotically), if
$$
\sqrt{D}\left(T_D-(q,\theta) \right) = \frac{1}{\sqrt{D}} \sum_{i=1}^{D} \tilde{I}_{(q,\theta)}^{-1} \tilde{\ell}_{(q,\theta)}(X_i^+) + o_P(1).
$$
\end{itemize}
\end{df}

Note that the typical asymptotics for estimator sequences rely on increasing sample size. The sample size in our problem is $D$, as the samples consist of the positive (observed) abundances $X_1^+,\cdots,X_D^+$. Thus, the sample size is a random quantity. Despite this, it is clear that as $N\to\infty$, we also have that $D\to\infty$ in probability, and we therefore think of the two asymptotic notions interchangeably.

One of the challenges is that in many models the efficient score is not amenable to be used in the same way as the ordinary score because the orthogonal projection $\varPi_F$ might not be available in closed form. In Proposition~\ref{Ef_Scor} (stated and proved in Section \ref{sec:effi}), we show that such a closed form can be obtained in our model, and give the expressions that ensue for the efficient score functions for estimating the parameters $\theta$ and $q$ in the model $\mathcal{P}^{+}.$

\subsection{Consistency and Efficiency}

In what follows, when the true model lies within the hypothesized class $\mathcal{P}$, we refer to the true parameters by $\theta_0$, $q_0$, and $F_0$, and to the true truncation by $\tau_0$. We first list some regularity assumptions that we have recourse to throughout.

\paragraph{Assumptions}
\begin{enumerate}
  \item \label{assume:compactness} [Compactness] $\varTheta$ is a compact subset of $\R^k$.
  
  \item \label{assume:identifiability} [Identifiability] The parameter $\theta$ is identifiable from the truncated density $S_{\theta}^{\tau}$, as defined by Equation \eqref{GenTruncated}.
  \item \label{assume:smoothness} [Continuity] For all $x$ in $\{1,\ldots, \tau\},$ $\theta \mapsto R_{\theta}(x)$ is a continuous function of $\theta$, and  $R_{\theta}(x) \geq \delta > 0$ for all $\theta$ in $\varTheta$ and $x\leq\theta_0$.
\end{enumerate}

Let us now move to the main results of this section, the consistency and efficiency of $\hat{q}_\tau$ and $\hat{\theta}_\tau$ whenever $\tau \leq \tau_{0},$ and some further properties that give more insight into these estimators. We begin with the consistency result stated below as Theorem \ref{Consistency}. 


\begin{theo} \label{Consistency} Under Assumptions \ref{assume:compactness}-\ref{assume:smoothness}, as $N$ tends to infinity, the following results hold:
 \begin{itemize}
  \item[$(i)$] If $\tau \leq \tau_{0},$  then $\hat{\theta}_\tau$ and $\hat{q}_\tau$ converge in probability to $\theta_{0}$ and $q_0$ respectively;
   \item[$(ii)$]  If $\tau > \tau_{0},$  then $\hat{\theta}_\tau$ converges in probability to the set of maximizers of $M^{\tau}(\theta) = \sum_{x=1}^{\tau}f^{+}(x)\log S_{\theta}^{\tau}(x) $.
 \end{itemize}
\end{theo}

The results in Theorem~\ref{Consistency} are remarkable since they ensure the consistency  of  $\hat{\theta}_\tau$ and $\hat{q}_\tau$ for a fixed $\tau$, as long as it is  smaller than or equal to its true value $\tau_0$ and identifiability holds. If, however, one chooses $\tau$ greater than $\tau_0,$ then the proposed estimators may not be consistent. (This leads to the challenge of choosing $\tau$ via model selection when $\tau_0$ is unknown, as described in Section \ref{sec:model-selection}). We now complement this consistency result with efficiency properties.

\begin{theo} \label{TheoEfficiency}
Consider Assumptions \ref{assume:compactness}-\ref{assume:smoothness}, and assume further that $\theta \mapsto R_{\theta}(x)$ is $\mathcal{C}^{2}$, that $\theta_0 \in \varTheta^\circ$, and that the efficient Fisher information is non-singular at $(q_0,\theta_0)$. Then,
$(\hat{q}_{\tau_0},\hat{\theta}_{\tau_0})$  is asymptotically efficient at $(q_0,\theta_0).$
\end{theo}

\begin{remark} \label{rem:dependence}
The  estimators  $\hat{q}_\tau$ and $\hat{\theta}_\tau$ have the following properties:
 \begin{itemize}
 	\item[$(i)$] $\hat{q}_\tau$ depends on the observations $x_i$ no greater than $\tau$ and on the cardinality of those $x_i$ that are greater than $\tau.$
 	
 	\item[$(ii)$] $\hat{\theta}_\tau$ depends only on the observations $x_i$ no greater than $ \tau.$
 \end{itemize}
These follow either from direct inspection or from the proof of Lemma \ref{lemma:score-equations} in the Appendix. In particular, $\hat{\theta}$ solves the efficient score equation~\eqref{EqTheta} which depends only on abundances $x_i$ no greater than $\tau.$ Now, from equation~\eqref{mleq}, for a given estimator $\hat{\theta}$ of $\theta,$ the estimator  $\hat{q}(\hat{\theta})$ depends on the $x_i$ greater than $\tau$ only through their cardinal $D-D_{\tau}$ and the property follows.
\end{remark}

Theorem~\ref{TheoEfficiency}, asserts the efficiency of $\hat{\theta}_\tau$ and $ \hat{q}_\tau,$ and through them of the corresponding estimator of the total number of species $\hat N_\tau$. Remark \ref{rem:dependence} also sheds light on the fact that the latter depends only on: (1) the threshold $\tau,$ (2) the number of observed species $D$ and (3) on the abundances of rare species (those that are not greater than $\tau$). In other words, as in the case of pure truncation, the abundant species contribute only through their cardinality. That said, $\hat N_\tau$ distinguishes itself by using this cardinality to estimate how to weigh appropriately the respective contributions of both the rare and abundant species, using the parameter $q$.

\section{Simulations and Experiments} \label{sec:simulations-experiments}

To illustrate the impact of truncation on our ability to estimate the number of species, we give some numerical simulations and experiments. To make our theoretical work concrete and results easily reproducible, we consider simple parametric families. In particular we look at a single Poisson distribution and a Gamma-Poisson mixture, which gives rise to the negative binomial distribution. In Section \ref{sec:simulations}, we perform synthetic experiments for both, and use this to illustrate the heuristic method of selecting the best truncation. In Section \ref{sec:experiments}, we consider real data in the form of literary texts, and confine ourselves to the negative binomial model. In order to be able to compare to a known ground truth, we adapt our number of species framework to the very related observational richness problem, and show that the choice of truncation has a significant impact on estimation accuracy.

\subsection{Number of Species Simulations} \label{sec:simulations} 

\subsubsection{Algorithms to compute $\hat{\theta}_{\tau}$}

As we take $R_\theta$ to be a parametric family, many of the EM-style MLE algorithms for parameter estimation in such frameworks can be adapted to zero- to $\tau$-truncated versions of the distributions. This is all that's needed since, for a fixed value of $\tau,$ computing $\hat{\theta}_{\tau}$ is an optimization problem that amounts to solving the efficient score equations in \eqref{Def_CSPE12}, and by Theorem~\ref{Consistency} this is equivalent to solving equation \eqref{EqTheta}. For example, when $R_\theta$ is a Poisson distribution, it is not difficult to check that the truncated MLE \eqref{EqTheta} becomes exactly
 \begin{equation} \label{eq:fixed-point}
  \bar{X}^{\tau}- \sum_{x=1}^{\tau}xS_{\theta}^{\tau}(x) = 0, \text{ with } 
  \bar{X}^{\tau} = \frac{1}{D_{\tau}}\sum_{i=1}^{D_{\tau}}x_i
 \end{equation}
leading to the fixed point equation
 $\theta =  \bar{X}^{\tau} \frac{\textbf{P}_{\theta}(\tau)-\exp(-\theta)}{\textbf{P}_{\theta}(\tau-1)}$
in which $\textbf{P}_{\theta}$ stands for the cumulative distribution function of the Poisson model with parameter $\theta.$ This is equivalent to moment-matching and the solution $\hat{\theta}_{\tau}$ could be found numerically by performing a bisection search, for example. Similar parameter searches can be performed for the truncated negative binomial distribution that we consider in this section. In a more complex model where $R_{\theta}$ is a finite mixture of Poisson distributions, that is when $R_{\theta}(x)= \sum_{j=1}^{J}\pi_jR_{\theta_j}(x)$ for all $x \geq 0,$ we can derive an EM algorithm for the truncated MLE similarly to the classical Poisson mixture. We do not elaborate this further, except to mention that each EM iteration entails the solution of fixed point equations, as in \eqref{eq:fixed-point}, for each Poisson component.

\paragraph{Design}
To investigate the performance of the new estimator and compare it to other existing estimators, we conducted a set of experiments with synthetic data.

In the first set of these experiments, we take the abundances of rare species to be distributed according to a single Poisson distribution with parameter $\theta$ and the nuisance distribution (of abundant species) is the uniform distribution on $\tau^{*},\ldots,\tau_{max}.$ The resulting distribution has density $ qR_{\theta}(x) + (1-q)U(x)$ with $0< q<1$ and $U$ the aforementioned uniform distribution. Now, for any fixed $N \in \{200,1000,5000,10000\},$ we generate a sample of size $N$ from the Bernoulli model with parameter $q \in \{0.4,0.6,0.8\},$ then generate the corresponding counts observations according to the Poisson or uniform model. The parameters $\tau^{*}$ and $\tau_{max}$ are fixed equal $10$ and $40$ respectively whereas $\theta$ ranges over $ \{0.6,1,1.5\}.$ The observed zero-truncated counts are used to compute our new estimator $\hat{N}_{\hat{\tau}}$ and some other existing estimators with which $\hat{N}_{\hat{\tau}}$ will be compared.

To show that the results extend to other parametric families, we perform a limited set second experiments, where we take the abundance of rare species to be distributed according to a Gamma-Poisson mixture, which leads naturally to the negative binomial distribution. In particular, in this case $\theta$ is two-dimensional, consisting of real parameters $r>0$ and $s>0$, and in Equation \eqref{eq:poisson-mixture} we have $\nu_\theta=\Gamma(r,s)$. This results in $R_\theta$ being the negative binomial distribution with parameters $r$ and $p=1/(1+s)$. We fix $p=0.8$ and take $r$ to vary over the range $\{0.5,1,2\}$. Larger values of $N$ are needed to learn this model even in the absence of nonparametric noise. We consider the range of $N \in \{10,000,20,000,50,000\},$ and we generate a sample of size $N$ from the Bernoulli model with parameter $q \in \{0.4,0.6,0.8\},$ then generate the corresponding counts observations according to the negative binomial or uniform model. That is, the observational model is as before, $ qR_{\theta}(x) + (1-q)U(x)$.

\begin{figure}[!ht]
\centering
\includegraphics[height=8cm, width=12cm]{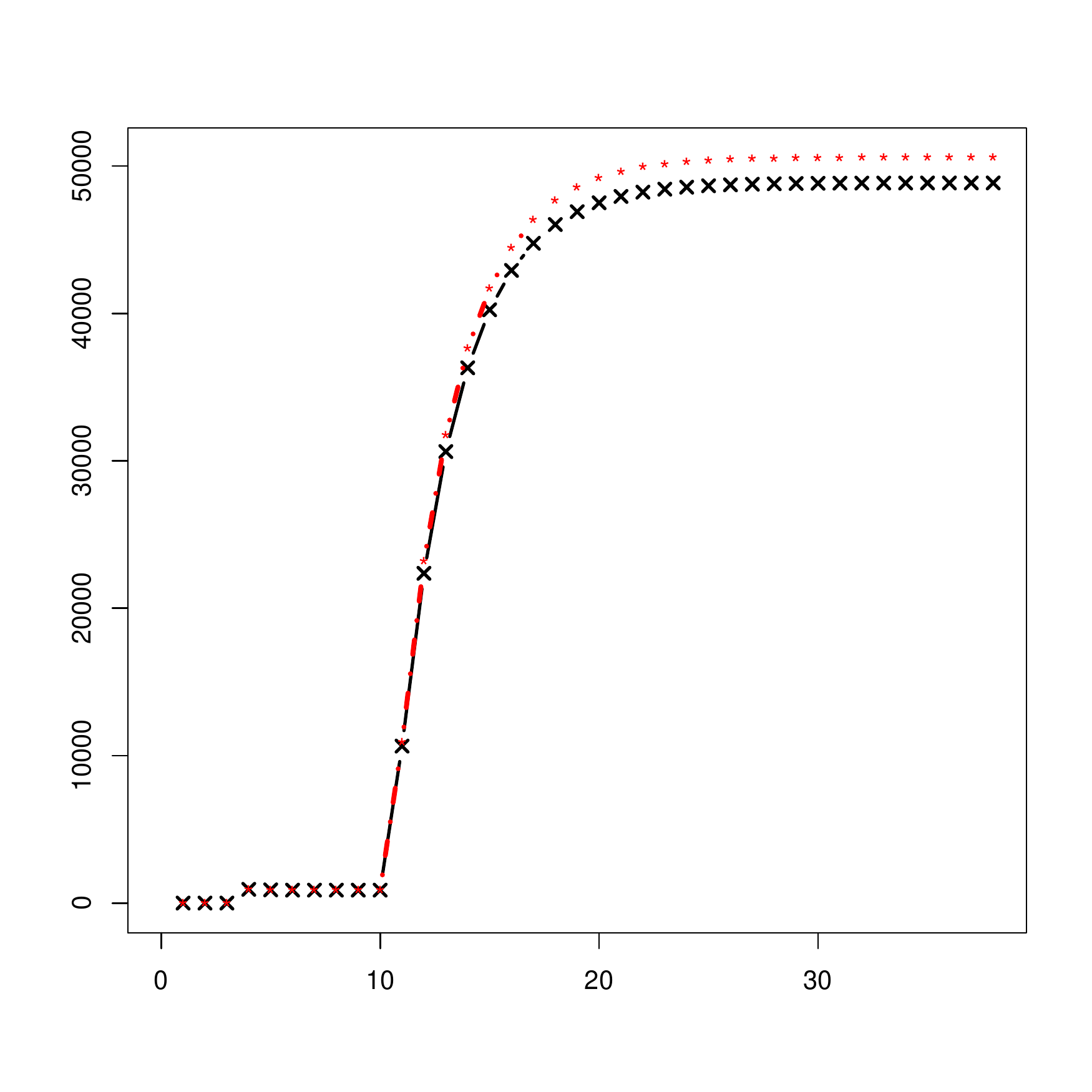}
\caption{\label{Fig_RiskProxy} Estimated risk of $\hat{N}_{\tau}$ as a function of $\tau$ (Lines X, in black) and its proxy $B(\tau)+pen(\tau)$ (dotdash $*$, in red) from the G-L method} .
\end{figure}

 
\paragraph{Risk approximation using G-L method}
We use the simulations as an opportunity to illustrate the G-L method and show the quality of the risk estimation by the proposed proxy in the selection rule. As displayed in Figure~\ref{Fig_RiskProxy}, the proxy $\hbias_\tau+\hvar_\tau$ provides a good approximation of the true risk when $\hvar_\tau$ is estimated by a bootstrap procedure as in Equation \eqref{eq:bootvar}. Note that in this numerical example we calculate the risk, bias proxy, and variance proxy for $N$ instead of $P(0)$. The approximation is remarkably accurate especially in the region where the estimator $\hat{N}_{\tau}$ is asymptotically unbiased $($that is for $\tau \leq \tau^{*}).$ Its remains satisfactory, but not overly so, for some $\tau $ greater than $\tau^{*}$. This indicates that the bootstrap procedure is a good choice to estimate $\hvar_\tau.$ Note that Figure~\ref{Fig_RiskProxy} corresponds to the results of simulations of the single Poisson model with parameters $q=0.6,$ $\theta = 1,$ and $N=1000.$ We obtain similar results for all other parameter choices.

\paragraph{Performances of $\hat{N}_{\hat{\tau}}$}
We focus on the performance of $\hat{N}_{\hat{\tau}}$ by calculating its Monte-Carlo mean and the renormalized standard error $(\frac{S_{e}}{N})$ based on $1000$ samples. We also investigate the bootstrap-based confidence interval for $N$ by providing the estimated non-coverage probabilities
 $$Inf=  \frac{1}{1000}\sum_{j=1}^{1000}\textbf{1}_{[N<N_{inf}^{(j)}]}$$
 and 
 $$ Sup = \frac{1}{1000}\sum_{j=1}^{1000}\textbf{1}_{[N>N_{sup}^{(j)}]},$$
 
where $I^{(j)}=[N_{inf}^{(j)},N_{sup}^{(j)}]$ is the bootstrap-based confidence interval using the estimated model from the $j^{th}$ Monte-Carlo sample. For the single Poisson model, the results are summarized in Table~\ref{table:Perf_N-1}. It is clear that the renormalized $Se$ decreases  when $\theta$ grows and increases as $q$ becomes larger. As the small values of $\theta$ characterize small abundances and that a high value of $q$ means that there is a large number of rare species in community $($according to the simulated model$),$ the observed variation of  $Se$ suggests that a high number of rare species will be  estimated with larger variance. We can also notice that the $Se$ decreases with $N$ in all simulated configurations showing the accuracy of the method when $N$ becomes larger. As the large values of $N$ describe the asymptotic regime of the estimators $\hat{\theta}_{\tau}$ and $\hat{q}_{\tau},$ we believe that the observed accuracy is related to the asymptotic efficiency of those estimators which improves the variance and then the mean square error $($MSE$)$ of $\hat{N}_{\hat{\tau}}$ as will be seen later. Table~\ref{table:Perf_N-2} summarizes the results for the Gamma-Poisson mixture model, with very comparable observations. Note that both sets of experiments show that we cannot rely on bootstrap confidence intervals as true intervals for the estimator. While the bootstrap is adequate in estimating the variability of the estimator, it does not accurately convey its location. It exhibits a clear skew to smaller values, which could be explained by the fact that resampling from the base distribution reduces the number of distinct observations. Therefore more principled methods are needed to go beyond point estimates in species richness estimation. One such avenue is through the use of concentration inequalities, \cite{benhamou17}.

 \begin{table}[!ht]
 \centering
  \caption{Performance of $\hat{N}_{\hat \tau}$ for single Poisson distributions. Inf and Sup are given in percentage ($\%$). \label{table:Perf_N-1}} 
  \begin{footnotesize}
  \begin{tabular}{rrrrrrrrrrrrrrrr}
          &	 
	&\multicolumn{4}{c}{$\theta=0.6$}&&\multicolumn{4}{c}{$\theta= 1$}&&\multicolumn{4}{c}{$\theta= 1.5$} \\
	\cline{3-6} \cline{8-11} \cline{13-16} 
 $q$ &  $N$ &    Mean  &   $ \frac{Se}{N}$& Inf ($\%$)& Sup ($\%$)&& Mean  &  $ \frac{Se}{N}$& Inf &Sup && Mean &   $ \frac{Se}{N}$& Inf & Sup \\
 \hline
$0.4$ &$200$ & $192$ &$0.116$& $1.5$ &$26.3$&& $200$ &$0.058$ & $2.5$&  $7.2$ &&  $199$  &$0.036$ &$2.2$ &$11.7$\\
 & $1000$ & $1005$ & $0.043$ &$2.9$  &$3.5$ &&$1001$&$0.024$& $3.6$ & $4.6$ && $1000$  &$0.014$ & $3.1$  &$4.2$\\
  &$5000$ &  $5003$ &$0.018$ &$3.0$& $3.4$ &&   $4999$ &$0.011$ & $3.0$ & $6.6$  &&   $5001$ &$0.006$   & $3.3$  &$3.7$\\
 &$10000$ & $10002$ &$0.013$& $3.5$& $4.4$ &&  $10002$ &$0.007$  & $3.3$ & $4.3$ && $10002$ &$0.005$   & $3.4$ &$4.6$\\
  &  &   & &  &   &&    &  &   &   &&   &   &   & \\
 $0.6$ &$200$  & $199$  &$0.133$   & $1.8$ & $11.7$ &&  $199$ &$0.073$   & $3.1$ &$9.1$ && $198$ & $0.042$ &  $ 2.0$ &  $12.7$\\
  &$1000$ & $1003$ & $0.055$   & $3.3$& $5.0$ && $1001$ &$0.030$  &  $3.9$ & $4.1$  &&$1000$ &$0.017$ & $2.9$  &  $2.7$\\
  &$5000$ & $5003$& $0.023$  &  $4.1$ &$3.5$ && $5001$ &$0.013$   & $3.5$  & $2.8$   && $5000$& $0.008$ & $2.7$ & $4.3$\\
 &$10000$& $10009$ &$0.017$   & $4.3$ &$3.7$ && $10005$& $0.009$ &   $4.0$   & $3.9$ &&  $9999$ &$0.006$ & $3.3$  &$4.0$\\ 
  &  &   & &  &   &&    &  &   &   &&   &   &   & \\
$0.8$  & $200$ & $192$& $0.160$ &$2.5$ &$15.5$    &&  $195$ & $0.079$ & $1.5$& $13.3$   && $196$ & $0.048$ & $1.1$& $17.0$\\
 & $1000$ & $1005$ & $0.063$ &$4.2$  &$5.0$ &&   $1002$ & $0.034$ &$3.7$ &$4.7$  && $999$ &$0.021$ & $3.5$ &$6.5$\\
 & $5000$ & $5017$ &$0.027$ & $5.2$ & $3.6$  &&  $5000$ & $0.015$ & $3.9$ & $4.0$ &&$4997$ &$0.009$ & $3.1$  &$4.1$\\
& $10000$& $10001$ &$0.019$ & $2.9$ & $4.6$  && $9999$& $0.011$ & $3.3$  &$4.6$ &&$9998$& $0.006$ & $3.2$ & $4.4$\\
\hline
  \end{tabular}
  \end{footnotesize}
 \end{table}

 \begin{table}[!ht]
 \centering
  \caption{Performance of $\hat{N}_{\hat \tau}$ for Gamma-Poisson mixtures ($p=0.8$). Inf and Sup are given in percentage ($\%$).\label{table:Perf_N-2}}
  \begin{footnotesize}
  \begin{tabular}{rrrrrrrrrrrrrrrr}
          &
        &\multicolumn{4}{c}{$r=0.5$}&&\multicolumn{4}{c}{$r= 1$}&&\multicolumn{4}{c}{$r=2$} \\
        \cline{3-6} \cline{8-11} \cline{13-16}
$q$     &$N$      &    Mean  &   $ \frac{Se}{N}$& Inf & Sup && Mean  &  $ \frac{Se}{N}$& Inf &Sup && Mean &   $ \frac{Se}{N}$& Inf& Sup \\
 \hline
$0.4$   &$10,000$ & $9,854$  & $0.038$ & $1.3$ & $10.4$ && $9,955$  & $0.012$ & $2.2$ & $14.6$ && $9,998$  & $0.003$ & $2.9$ & $9.0$  \\
        &$20,000$ & $19,413$ & $0.020$ & $0.0$ & $24.0$ && $19,867$ & $0.008$ & $1.1$ & $25.0$ && $19,981$ & $0.002$ & $1.3$ & $15.5$ \\
        &$50,000$ & $48,359$ & $0.013$ & $0.0$ & $64.0$ && $49,561$ & $0.005$ & $0.0$ & $50.0$ && $49,933$ & $0.001$ & $0.0$ & $39.0$ \\
        &         &          &         &       &        &&          &         &       &        &&          &         &       &        \\
$0.6$   &$10,000$ & $9,618$  & $0.042$ & $0.4$ & $25.0$ && $9,883$  & $0.015$ & $0.8$ & $16.3$ && $9,986$  & $0.003$ & $1.3$ & $13.3$ \\
        &$20,000$ & $19,222$ & $0.035$ & $0.4$ & $35.4$ && $19,823$ & $0.011$ & $1.1$ & $27.0$ && $19,964$ & $0.002$ & $1.3$ & $27.9$ \\
        &$50,000$ & $47,792$ & $0.018$ & $0.0$ & $71.0$ && $49,319$ & $0.005$ & $0.0$ & $72.0$ && $49,885$ & $0.002$ & $0.4$ & $53.2$ \\
        &         &          &         &       &        &&          &         &       &        &&          &         &       &        \\
$0.8$   &$10,000$ & $9,561$  & $0.053$ & $0.7$ & $23.1$ && $9,843$  & $0.016$ & $0.3$ & $27.0$ && $9,973$  & $0.004$ & $0.7$ & $23.3$ \\
        &$20,000$ & $18,770$ & $0.031$ & $0.0$ & $49.0$ && $19,623$ & $0.011$ & $0.1$ & $50.7$ && $19,968$ & $0.003$ & $3.2$ & $21.4$ \\
        &$50,000$ & $46,812$ & $0.019$ & $0.0$ & $86.0$ && $49,128$ & $0.006$ & $0.0$ & $76.0$ && $49,816$ & $0.002$ & $0.0$ & $80.0$ \\
\hline
  \end{tabular}
  \end{footnotesize}
 \end{table}

\paragraph{Comparison with other estimators}
We end the simulations by comparing the proposed estimator of the number of species to other existing one in literature. We focus entirely on the single Poisson model, which represents the ground truth 
assumption of many of theses estimators. We consider the Chao's estimator $\hat{N}_{Ch_0}$ defined as lower bound for $N$ and proposed in~\cite{Ch84}, the coverage based estimator $(\hat{N}_{CL})$ proposed
in~\cite{ChL92} by Chao and Lee, the estimator $\hat{N}_{CB}$ of $N$ using the expected proportion of duplicate species in the sample (by Chao and Bunge in~\cite{ChB02}), the nonparametric MLE $\hat{N}_{WL_0}$ 
of $N$ using a penalized likelihood (by Wang and Lindsay in~\cite{WL05}) and $\hat{N}_{LB}:$ an extension of Chao's estimator proposed by Lanutheang and B\"{o}hning in~\cite{LB11}. The criteria used for this 
comparison (Mean, rMAE: relative Mean Absolute Error and rMSE: relative Mean Square Error) are computed and presented in Table~\ref{table:Comp_Est}. The six estimators display a good performance in all 
simulated configurations and $\hat{N}_{\hat{\tau}}$ seems to better estimate $N$ than all other methods. This is quantified in Table~\ref{table:Comp_Est}, by the remarkably small value of $rMSE$ as compared to 
the others. This shows that, despite our results being about the asymptotic efficiency of $\hat{\theta}_{\tau}$ and $\hat{q}_{\tau}$, we can expect finite-sample improvements for the estimator
$\hat{N}_{\hat{\tau}}$, when $N$ is moderately large. Also note that all six estimators become less reliable for very small value of $\theta$ or large value of $q$ explaining thus the common difficulty for 
these approaches to better approximate $N$ in the case of a large number of rares species, which touches upon the inherent problems of unidentifiability~\cite{ML07}.

 \begin{table}[!ht]
 \centering
  \caption{Comparison of $\hat{N}_{\hat \tau}$ with five other estimators of $N$ using $1000$ monte-carlo samples. $\hat{N}_{Ch_0}:$ Chao's estimator as lower bound on $N$ in~\cite{Ch84};
  $\hat{N}_{CL}:$ The coverage based estimator of $N$ by Chao and Lee in~\cite{ChL92};$\hat{N}_{CB}:$ Estimator of $N$
  using the expected proportion of duplicate species in the sample (by Chao and Bunge in~\cite{ChB02}); 
  $\hat{N}_{WL_0}$ Nonparametric MLE of $N$ using a penalized likelihood (by Wang and Lindsay in~\cite{WL05}) and 
  $\hat{N}_{LB}$ is an extension of Chao's estimator proposed by Lanutheang and B\"{o}hning in~\cite{LB11}. \label{table:Comp_Est}}
  \begin{footnotesize}
  \begin{tabular}{rrrrrrrrrrrrr}
          &	 
	&\multicolumn{3}{c}{$\theta=0.6$}&&\multicolumn{3}{c}{$\theta= 1$}&&\multicolumn{3}{c}{$\theta= 1.5$} \\	
	\cline{3-5} \cline{7-9} \cline{11-13} 
  \hspace*{1cm}
  $q$ &Est &    Mean  &   rMAE& rMSE&& Mean  &  rMAE& rMSE&& Mean &  rMAE& rMSE \\
 \hline
$0.4$& $\hat{N}_{\hat{\tau}}$ &$1005$& $0.034$ &$ 0.185$&& $1001$& $0.019$& $0.058 $&&$1000$& $0.011$& $0.019$\\
&$\hat{N}_{Ch_0}$   &   $1010$ &$0.045$ & $0.341$&&$ 1002$ &$0.026$&$ 0.108$&& $1001$ &$0.016$& $0.041$\\
&$\hat{N}_{CL}$ &  $1015$ &$0.040$ & $0.28$6&&$ 1007$& $0.023$& $0.084$&& $1004$& $0.013$& $0.029$\\
& $\hat{N}_{CB}$ &   $1054$ &$0.132$& $23.651$&& $1004$& $0.035 $&$0.227$&& $1002$& $0.018$ &$0.051$\\
& $\hat{N}_{WL_0}$ &  $1041$&$ 0.058$& $ 0.731$&&$ 1024$&$0.035$& $0.292$ &&$1017$ &$0.023$ &$0.146$\\
& $\hat{N}_{LB}$  &  $1026$ &$0.092$& $ 1.717$&&$ 1022$& $0.046$& $0.482$&& $1014$ &$0.028$& $0.162$\\ 
&    &    &  &   &&      &   &   &&      &   & \\
$0.6$ &$\hat{N}_{\hat{\tau}}$& $1003$ &$0.043$&  $0.298$&& $1001$ &$0.024$& $0.088$ &&$1000$&$ 0.014$& $0.029$\\
& $\hat{N}_{Ch_0}$   &  $1007$& $0.056$&  $0.522$&&$ 1003$&$0.031$& $0.160$ &&$1002$& $0.020$& $0.065$\\
&$\hat{N}_{CL}$  & $1015$ &$0.051$&  $0.434$&& $1008$& $0.027$&$ 0.125$&& $1005$&$ 0.017$ &$0.045$\\
&$\hat{N}_{CB}$ &   $1037$&$0.119$ & $3.956$&& $1005$ &$0.043$& $0.315$ &&$1002$ &$0.022$& $0.080$\\
&$\hat{N}_{WL_0}$ &   $1044$& $0.072$ & $1.122$&& $1034$& $0.047$& $0.510$&& $1025$&$ 0.032$& $0.288$\\
&$\hat{N}_{LB}$ &    $1045$& $0.113$ & $2.789$&& $1031$ &$0.057$& $0.704$&& $1018$& $0.034$& $0.250$\\ 
&    &    &  &   &&      &   &   &&     &   & \\
$0.8$&$\hat{N}_{\hat{\tau}}$ &$1005$& $0.051 $ &$0.401 $&&$1002$& $0.027$& $0.118 $&& $999$ &$0.017$ &$0.045$\\
&$\hat{N}_{Ch_0}$     &$1009$& $0.062$ & $0.621$&& $1006$& $0.037$& $0.218$&& $1003$ &$0.023 $&$0.088$\\
& $\hat{N}_{CL}$  &$1020$ &$0.058$ &$ 0.553$&& $1011$& $0.032$ &$0.169$&& $1006$ &$0.020 $&$0.065$\\
&$\hat{N}_{CB}$   &$1038$& $0.128$ & $3.719$&& $1007$& $0.051$& $0.433$&& $1003$ &$0.026$& $0.111$\\
&$\hat{N}_{WL_0}$   & $1062$& $0.088$  &$1.550$&& $1046$&$0.060$& $0.835$&& $1031$ &$0.040$& $0.405$\\
& $\hat{N}_{LB}$  & $1059$ &$0.126$ & $3.452$&& $1041$& $0.069 $&$1.054$&&$ 1019$& $0.038$& $0.294$\\
\hline
  \end{tabular}
  \end{footnotesize}  
 \end{table} 
  
\subsection{Observational richness in text data}  \label{sec:experiments}

Rather than estimating the absolute number of species, an important extension of the species richness problem is concerned with estimating the number of distinct species to be observed in a sample larger than the current sample of individuals. Indeed, the abundance data $X_1^+,\cdots,X_D^+$ is ostensibly obtained by performing a sampling of individuals. If the said sample is enlarged, then how do the new abundances relate to the original ones? In the words of Fisher \cite{Fisher1943}, in a pure Poisson abundance model: ``Obviously, [the parameter $\lambda$] will be proportional to the size of the sample taken [...]''. This is most easily seen in the individual sampling model of Equation \eqref{eq:binomial}: when the binomial size parameter is changed from $n$ to $n'=\gamma n$, the parameters of the corresponding Poisson mixture are changed from $\lambda=np_j$ to $\lambda'=n'p_j=\gamma \lambda$.

Generally in a Poisson mixture model, therefore, a $\gamma$ factor increase in the sample size is equivalent to a $\gamma$ dilation of the mixture distribution. Let $\mathbf{E}^\gamma[D]$ denote the expected number of distinct symbols in the enlarged sample, and thus $\mathbf{E}^1[D]=\mathbf{E}[D]=N(1-qR_\theta(0))$. The observational richness estimation problem can thus be concretely stated as the problem of estimating $\mathbf{E}^\gamma[D]$, based on $X_1^+,\cdots,X_D^+$.

One application of the observational richness problem is to forecast the vocabulary of an author, from a portion of their text. This was popularized in the work of \cite{EfronThisted1976}, who applied this methodology to the complete works of William Shakespeare. The problem goes back to the work of \cite{GoodToulmin1956}, who approached it from an empirical Bayesian perspective, without any specific parametrization. The earlier work of \cite{Fisher1943} also implicitly addressed the same problem.

Here, we restrict ourselves to the context of a parametric Poisson mixture abundance model for $R_\theta$, that is as in Equation \eqref{eq:poisson-mixture}, with $\nu=\nu_\theta$ appropriately parametrized by $\theta$. We require the family of such densities to be closed under dilation, as in for all $\theta$ and $\gamma>0$, there exists $\theta^\gamma$, such that for all measurable subsets $A$, $\nu_{\theta^\gamma}(A)=\nu_\theta(A/\gamma)$. Furthermore, we assume that for fixed $\gamma$, the transformation $\theta \mapsto \theta^\gamma$ is continuous in the sense that if a sequence $\theta_i \to \theta$ then the sequence $\theta^\gamma_i \to \theta^\gamma$. Note that for discrete mixtures, the scaling simply shifts the supports by $\gamma$, and for continuous mixtures it expands and scales the density by $\gamma$, and the requirement in either case is for the resulting density to remain an element of the parametric family.

As we focus primarily on text data, the Gamma-Poisson mixture family is very well-suited. Recall that in this case $\theta$ is two-dimensional, consisting of real parameters $r>0$ and $s>0$, and $\nu_\theta=\Gamma(r,s)$, and the corresponding negative binomial distribution with parameters $r$ and $p=1/(1+s)$. To dilate the Gamma distribution, it is easy to see that one simply scales $s'=\gamma s$. This corresponds to a transformation of the negative binomial parameter $p'=1/(1+\gamma(1-p)/p)$.

This paper's framework applies to this problem as follows. If the rare abundances are well modeled by a Gamma-Poisson mixture while the abundant ones are not, then our framework allows us to efficiently learn the parameters $q$ and $\theta$. By continuity, for fixed $\gamma$ we also have an efficient estimator of $\theta^\gamma$. Since $N$ is assumed to stay constant, we then have
$$
    N = \frac{\mathbf{E}[D]}{1-qR_\theta(0)} = \frac{\mathbf{E}^\gamma[D]}{1-qR_{\theta^\gamma}(0)}.
$$
We could therefore use our estimates $\hat q_\tau$ and $\hat \theta_\tau$ to evaluate  $\hat \theta_\tau^\gamma$ and thus to estimate $\mathbf{E}^\gamma[D]$ as follows:
$$
\widehat{\mathbf{E}^\gamma[D]}_\tau := \frac{1-\hat q_\tau R_{\hat \theta_\tau^\gamma}(0)}{1-\hat q_\tau R_{\hat \theta_\tau}(0)} D.
$$

\begin{figure}
\begin{center}
\includegraphics[scale=0.5]{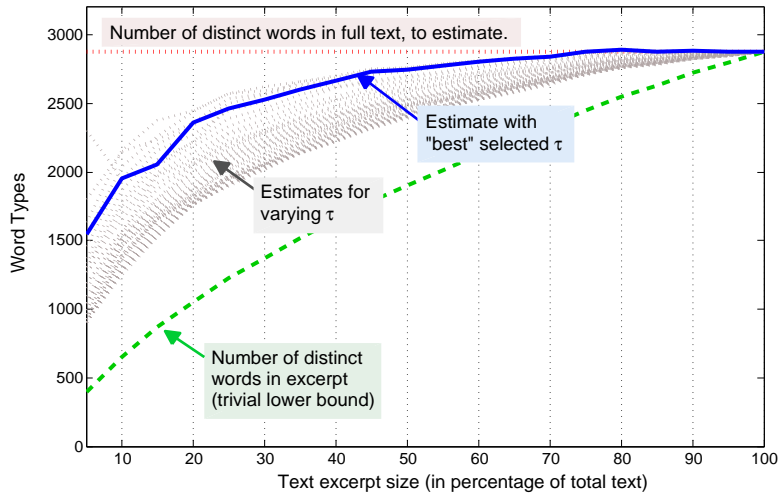}
\end{center}
\caption{Estimating the vocabulary growth in Moli\`ere's \emph{Tartuffe} play.}
\label{fig:moliere}
\end{figure}

The data we look at is French playwrite Moli\`ere's \emph{Tartuffe} play, which we gradually observe a portion of and try to estimate the number of distinct vocabulary words. Thus, the scale $\gamma$ is the ratio of the total text size to the size of the observed text, varying from $0$ to $100 \%$. For this problem, we illustrate $\widehat{\mathbf{E}^\gamma[D]}_\tau$ for various choices of $\tau$ and also the Goldenshluger-Lepski selected $\hat \tau$ in Figure \ref{fig:moliere}. Note how quickly the result becomes an accurate estimate of the vocabulary. But most importantly, note how sub-optimal choices of the truncation can adversely affect the performance of the estimator.

\section{Conclusion} \label{sec:conclusion}

 In this paper, we revisited the species richness estimation problem and studied a commonly followed practice of truncating the data into rare and abundant species. We proposed a semiparametric framework to model such a truncation as a parametric component well-suited to model rare species and a nonparametric nuisance component to cover the abundant species in an agnostic manner. We showed that asymptotic efficiency in this framework requires handling the truncation more delicately. This is in particular true if the rare species model has a significant overlap with the abundant species. Finally, we proposed a heuristic method to learn a good truncation threshold from data.

Several possible avenues of investigation may be proposed. We already mentioned the importance of going beyond point estimates. One would also like to relax the assumption that the abundant species are truly located entirely away from zero. In particular, it is important to handle the situation when such a dichotomy arises from an underlying binomial mixture model. Some recent approaches to species richness have successfully used Chebyshev polynomials as a fitting model, see for example \cite{orlitsky16}, and one would like to understand the relationship between such fits and mixtures of Poissons. Finally, one would hope that a truncation threshold that automatically conforms to the underlying model could make the most of the available data and thus give a fundamental theoretical edge, perhaps in the form of adaptive rates.

\appendix

\section{Proofs} \label{sec:proofs}

\subsection{Proof of Proposition~\ref{Equiv_MLE}\\}
For any fixed $\tau,$ the classical conditional MLE satisfies
$$
  \hat{N}_{\mathrm{classical}} = D_{a} + \frac{D_{\tau}}{1-R_{\hat{\theta}_{\tau}}(0) } =
 D+ \frac{D_{\tau}R_{\hat{\theta}_{\tau}}(0)}{1-R_{\hat{\theta}_{\tau}}(0)}.
$$
Let us consider now the new conditional MLE proposed in this work.
\begin{eqnarray*}
  \hat{N}_{\tau}
  &=&\frac{D}{1-\hat{q}_{\tau}R_{\hat{\theta}_{\tau}}(0)}\\
    &=& D +  \frac{D\hat{q}_{\tau}R_{\hat{\theta}_{\tau}}(0)}{1-\hat{q}_{\tau}R_{\hat{\theta}_{\tau}}(0)}.
\end{eqnarray*}
But using equation \eqref{eq:q-hat}, we have 
$$\frac{\hat{q}_{\tau}R_{\hat{\theta}_{\tau}}(0)}{1-\hat{q}_{\tau}R_{\hat{\theta}_{\tau}}(0)} = \frac{D_{\tau}R_{\hat{\theta}_{\tau}}(0)}{D \sum_{k=1}^{\tau}R_{\hat{\theta}_{\tau}}(k)}$$
from which we get
$$
  \hat{N}_{\tau}= D + \frac{D_{\tau}R_{\hat{\theta}_{\tau}}(0)}{\sum_{k=1}^{\tau}R_{\hat{\theta}_{\tau}}(k)}.
$$
Now,
if $R_{\theta}$ is supported on $\{0,\ldots,\tau\},$ then $\sum_{k=1}^{\tau}R_{\hat{\theta}_{\tau}}(k) $ equals
$ 1-R_{\hat{\theta}_{\tau}}(0)$ in the last expression which finally gives $ \hat{N}_{\tau}=  \hat{N}_{\mathrm{classical}} $. \cqfd

\subsection{Proof of Proposition~\ref{Equiv_MLE_Zelt}\\}
For $\tau$ equals $2$ and $R_{\theta}$ being the Poisson distribution with parameter $\theta,$ it is not difficult to see that $ \hat{\theta}_{\tau} =\hat{\theta}_\mathrm{Zelterman} = 2n_{2}/n_{1}.$ Then, \begin{eqnarray*}
  \hat{N}_{\tau}
    &=&   D + \frac{D_{\tau}R_{\hat{\theta}_{\tau}}(0)}{\sum_{k=1}^{\tau}R_{\hat{\theta}_{\tau}}(k) }\\
    &=& D + \frac{D_{\tau}}{\hat{\theta}_{\tau}+\hat{\theta}_{\tau}^{2}/2}  \\
    &=& D + n_{1}^{2}/2n_{2}.
\end{eqnarray*}
The last equality holds by replacing $ \hat{\theta}_{\tau}$ and $D_{\tau}$ by $2n_{2}/n_{1}$ and $n_{1}+n_{2}$ respectively.\cqfd

%


\subsection{Proof of Theorem~\ref{Consistency}\\}


To prove $(i),$ we use the fact that $\hat{\theta}$ is the maximum likelihood estimator in the model with density $S_{\theta}^{\tau}$. Note that maximizing the likelihood of equation \eqref{TruncVrais} amounts to maximizing in $\theta$ the criterion $\mathcal{L}_{D}(\theta) = \sum_{x=1}^{\tau} \frac{n_{x}}{D_{\tau}} \log S_{\theta}^{\tau}(x)$ which as $N$ tends to infinity converges almost surely  to  $ \mathcal{L}(\theta) =\sum_{x=1}^{\tau} S_{\theta_0}^{\tau}(x) \log S_{\theta}^{\tau}(x)$ when $ \tau \leq \tau_0$. Moreover, we have
$$
  |\mathcal{L}_{D}(\theta) - \mathcal{L}(\theta)|  \leq
  \sum_{x=1}^{\tau} \left|\frac{n_{x}}{D_{\tau}} -  S_{\theta_0}^{\tau}(x)\right|~
  |\log S_{\theta}^{\tau}(x)|.
$$
On the right hand side of this inequality, $\frac{n_{x}}{D_{\tau}} - S_{\theta_0}^{\tau}(x)$ converges almost surely, thus in probability, to zero. Also, $|\log S_{\theta}^{\tau}(x)|$ is bounded since $R_{\theta}(x) \geq \delta>0$, for all $\theta \in \varTheta$ and all $x \leq \tau.$ We conclude that
$$
  \sup_{\theta \in \varTheta} \vert \mathcal{L}_{D}(\theta) - \mathcal{L}(\theta) \vert \to 0 \quad\textrm{in probability.}
$$

It is easy to see that
$$
  \mathcal{L}(\theta) -\mathcal{L}(\theta_0) = \sum_{x=1}^{\tau} S_{\theta_0}^{\tau}(x) \log \frac{S_{\theta}^{\tau}(x)}{S_{\theta_0}^{\tau}(x)}
$$
attains uniquely its maximum $($equals zero$)$ at $\theta_0$ since the true model $S_{\theta_0}^{\tau}$ is identifiable, as assumed. We then obtain
\begin{equation} \label{cond2}
  \sup_{\theta:d(\theta, \theta_0)\geq \varepsilon} \mathcal{L}(\theta) < \mathcal{L}(\theta_0) ,
\end{equation}
where $d(\theta, \theta_0)$ is the euclidean distance between $\theta$ and $\theta_0$. As $\hat{\theta}$ maximizes $\mathcal{L}_{D},$ we have $\mathcal{L}_{D}(\hat{\theta}) \geq \mathcal{L}_{D}(\theta_0) - $o$_{\po}(1).$  This, together with the condition in equation \eqref{cond2} and the above convergence in probability, entails that $\hat{\theta}$ converges in probability to $\theta_0$ as $N$ tends to infinity. This result holds from Theorem $5.7$ in~\cite{AWV98}.

To end part $(i)$ of the theorem, recall Equation \eqref{eq:q-hat}:
$$
  \hat{q}(\theta) = \frac{1}{R_{\theta}(0)+ \frac{D}{D_{\tau}} \sum_{k=1}^{\tau} R_{\theta}(k)}.
$$  
We then observe  from the law of large numbers that as $N$ tends to infinity, $\frac{D}{D_{\tau}}$ converges almost surely to $\frac{1-q_{0}R_{\theta_0}(0)}{q_{0}\sum_{k=1}^{\tau} R_{\theta_0}(k)}$ when $ \tau \leq \tau_0.$ Recall that we assume $R_\theta$ to be continuous in $\theta$ for each $x$. Thus using the continuous map theorem and the convergence in probability of $\hat{\theta}$ to $\theta_0$, we find that $R_{\hat{\theta}}(0)$ and $ \sum_{k=1}^{\tau} R_{\hat{\theta}}(k)$ converge in probability to $R_{\theta_0}(0)$ and $\sum_{k=1}^{\tau} R_{\theta_0}(k)$ respectively when $\tau \leq \tau_0$. We finally obtain the convergence in probability of $\hat{q}(\hat{\theta})$ to
$$
  \hat{q}(\theta_0) = \frac{1}{R_{\theta_0}(0)+ \left\{ \frac{1-q_{0}R_{\theta_0}(0)}
  {q_{0}\sum_{k=1}^{\tau} R_{\theta_0}(k)}\right\} \sum_{k=1}^{\tau} R_{\theta_0}(k)} = q_0,
$$
using once again the continuous map theorem. This ends the proof of part $(i)$ of Theorem~\ref{Consistency}.
 
Similar arguments to what we have given here can be used to prove part $(ii)$ of the Theorem, namely that if  $\tau > \tau_0$, then $\hat{\theta}$ converges in probability to the set of maximizers of $M^{\tau}(\theta) = \sum_{x=1}^{\tau}f^{+}(x)\log S_{\theta}^{\tau}(x)$, in the sense that the probability of falling in an $\epsilon$-dilation of this set tends to $1$ as $N\to\infty$. \cqfd

\subsection{Efficient score functions and efficient Fisher information}
\label{sec:effi}

We now build up some notation.
Let $\mathcal{G}$ denote the set of measurable functions defined on the support of $F$ by
 \begin{equation}
  \mathcal{G}  = \left\{G,\ \sum_{x>\tau}F(x)G(x)=0 \text{ and }  \sum_{x>\tau}F(x)G^{2}(x)< \infty
  \right\}.
 \end{equation}
For a given $G$ in $\mathcal{G},$ a real number $a$ and a vector $b$ of dimension $k$ let us define $q_{t} = q + at,$ $\theta_{t} = \theta + bt$ and $F_{t} = F(1+tG).$ This parametrization of $F,$ $q$ and $\theta$ defines a path (a one-dimensional sub-model) $f_{t}^{+} = f_{(q_t,\theta_t,F_t)}^{+}$ in the model $\mathcal{P}^{+}.$ To simplify the notation, we let $f^{+}$ stand for $f_{(q,\theta,F)}^{+}$  and $f$ for $f_{(q,\theta,F)}$. Recall the definition of score functions.
 
\begin{df}
A differentiable path is a map $t\mapsto f_{t}^{+}$ from a neighborhood $[0,\varepsilon)$ of $0$ to $\mathcal{P}^{+}$ with $f_{0}^{+} = f^{+}$  such that, for some measurable real valued function $g,$ one has
\begin{equation}
  \sum \left( \frac{\sqrt{f_{t}^{+}} -\sqrt{f^{+}_{\phantom{t}}} }{t} -
  \frac{1}{2}g\cdot\sqrt{f_{\phantom{t}}^{+}} \right)^{2} \rightarrow 0 \mbox{ as } t \rightarrow 0.
\end{equation}
The one-dimensional sub-model $\left\{ f_{t}^{+}, t \in [0,\varepsilon)\right\}$ is then said to be \emph{differentiable in quadratic mean} at $f^{+}$ with \emph{score function} $g$.
\end{df}


A more useful way to determine the score function of a model such as $\left\{ f_{t}^{+}, t \in [0,\varepsilon)\right\}$ is to take the derivative with respect to $t$ of the log-likelihood at $t=0,$ that is
\begin{equation}
   g = \frac{\ud }{\ud t} \Bigg\vert_{t=0} \log f_{t}^{+}.
\end{equation}

We will use a dot-notation to indicate differentiation with respect to a parameter. Recall first the parametric score function $\dot{\ell}_q$ and the parametric vector score function $\dot{\ell}_\theta$ which are the partial derivative and gradient with respect to $q$ and $\theta$ respectively of the log-likelihood in the full model. We have respectively
\begin{equation}\label{Eq:scores}
 \dot{\ell}_{q} = \frac{R_{\theta}-F}{f} + \frac{R_{\theta}(0)}{1-qR_{\theta}(0)} \mbox{ and }
  \dot{\ell}_{\theta} = \frac{q\dot{R}_{\theta}}{f} + \frac{q\dot{R}_{\theta}(0)}
  {1-qR_{\theta}(0)},
\end{equation}
with $\dot{R}_{\theta}$ the gradient function of the density $ R_{\theta}.$

In the model defined in Equation \eqref{Eq:Model_t}, a straightforward calculation shows that the score function $g$ of the one-dimensional sub-model is such that
$$
g = a\dot{\ell}_q + \langle b,\dot{\ell}_\theta\rangle  + \frac{(1-q)FG}{f}
$$
where $a$ and $b$ are the scaling scalar and $k$-dimensional vector of the parametrizations $q_t$ and $\theta_t$ respectively, and where $\langle \cdot,\cdot\rangle$ denotes the usual inner product.

Now, we recall briefly the notions of tangent set and efficient score function for the model considered here. The maximal tangent set to the model $\mathcal{P}^{+} $ at $f^{+}$ is the set of all score functions of a one-dimensional sub-model. We denote it $\mathcal{\dot{P}}^{+}$, and in our case it is given by
 \begin{equation}\label{Eq:esp_tan}
   \mathcal{\dot{P}}^{+} = \left\{ g =  a\dot{\ell}_q + \langle b,\dot{\ell}_\theta\rangle
   + \frac{(1-q)FG}{f};
   u \in \R^{k+1} \mbox{, }  \mbox{ and  } G \in \mathcal{G} \right\}.  
 \end{equation}
Consider again the path $t\mapsto f_{(q,\theta,F_t)}^{+}$ related to the model $\mathcal{P}^{+} $, but now with the parameters $q$ and $\theta$ fixed, then the tangent set at $f^{+}$ for the nonparametric part of the model in \eqref{Eq:Model_t} is denoted and given by
 \begin{equation}\label{Eq:esp_tan0}
  \mathcal{\dot{P}}_{F}^{+} = \left\{ h =\frac{(1-q)FG}{f}, 
  \text{ with } G \in \mathcal{G}\right\}.
 \end{equation}
 
The \emph{efficient score function} related to a given component $\alpha$ of the parameter vector $(q,\theta_1,\cdots,\theta_k)$ is then defined component-wise as $\tilde{\ell}_{\alpha}=\dot{\ell}_{\alpha} - \varPi_{F} \dot{\ell}_{\alpha},$ where $\varPi_{F}$ is the orthogonal projection onto the closure of the linear space spanned by $\mathcal{\dot{P}}_{F}^{+}.$

The expressions of the efficient score functions are given in the following proposition, the coefficients of the efficient Fisher information matrix are displayed in the proof of this proposition.

 \begin{prop} \label{Ef_Scor}
  The efficient score functions for estimating the parameters $q$ and $\theta$ are given for $x \geq1 $ by
  \begin{equation} \label{Ef_Scorq}
   \tilde{\ell}_{q}(x) = \frac{1}{q} \ind\{x\leq \tau\} -
   \frac{\sum_{k=0}^{\tau}R_{\theta}(k)}{1-q\sum_{k=0}^{\tau}R_{\theta}(k)} \ind\{x> \tau\}+
   \frac{R_{\theta}(0)}{1-qR_{\theta}(0)}
  \end{equation}
  and
  \begin{equation}\label{Ef_ScorTheta}
   \tilde{\ell}_{\theta}(x) = \frac{\dot{R}_{\theta}(x) }{R_{\theta}(x)} \ind\{x\leq \tau\}-
   \frac{q\sum_{k=0}^{\tau}\dot{R}_{\theta}(k)}{1-q\sum_{k=0}^{\tau}R_{\theta}(k)} \ind\{x> \tau\}+
   \frac{q\dot{R}_{\theta}(0)}{1-qR_{\theta}(0)}
  \end{equation}
  respectively. The efficient Fisher information $\tilde{I}$ is a matrix of order $(k+1)$ with coefficients
  given by equations \eqref{Iqtilde}-\eqref{Iqthetatilde}.
\end{prop}
\paragraph{Proof}

Recall from \eqref{Eq:esp_tan0} the definition of the tangent set of the nonparametric part of the one-dimensional sub-model:
$$
  \mathcal{\dot{P}}_{F}^{+} = \left\{ h =\frac{(1-q)FG}{f},
  \text{ with } G \in \mathcal{G}\right\},
$$
and let $\overline{lin}(\mathcal{\dot{P}}_{F}^{+})$ denote the closure of the linear space spanned by $\mathcal{\dot{P}}_{F}^{+}$ in $\LL^{2}(f^{+}).$ With this notation, recall that the efficient score function related to a given component $\alpha$ of the parameter vector $(q,\theta_1,\cdots,\theta_k)$ is defined component-wise as $\tilde{\ell}_{\alpha}=\dot{\ell}_{\alpha} - \varPi_{F} \dot{\ell}_{\alpha},$ where $\varPi_{F}$ is the orthogonal projection onto $\overline{lin}(\mathcal{\dot{P}}_{F}^{+})$ in $\LL^{2}(f^{+})$.

To reduce clutter, let $\dot\ell$ refer to a particular component $\dot\ell_\alpha$. We first give a closed form expression of the orthogonal projection. In particular, we have that:
\begin{equation} \label{eq:proj}
 (\varPi_{F}\dot\ell)(x) = \begin{cases}
                      0 \text{ if } 1\leq x \leq \tau, \\
                      \dot\ell(x) - c(\dot\ell) \text{ if } x > \tau,
                     \end{cases}
\end{equation}
where $c(\dot\ell)$ is a constant depending on $\dot\ell$ as follow
\begin{equation}\label{eq:constant}
  c(\dot\ell) = \frac{\sum_{x>\tau}\dot\ell(x)f(x) }{\sum_{x>\tau}f(x)}.
\end{equation}

To see this, first observe that for every score function $\dot\ell$ in the model $\mathcal{P}^{+},$ the projection $\varPi_{F}\dot\ell$ is an element of the subspace $\overline{lin}(\mathcal{\dot{P}}_{F}^{+})$ so that it must be a linear combination (or a limit thereof) of elements of the form $\frac{(1-q)FG_0}{f}$ for some $G_0 \in \mathcal{G}$. Since the latter all vanish on the set $\{1,\ldots,\tau\}$, so does $\varPi_{F}\dot\ell$.

Next, let $\tilde{h}$ be any $\LL^{2}(f^{+})$-integrable function that is orthogonal to the space $\overline{lin}(\mathcal{\dot{P}}_{F}^{+})$, that is:
$$
  \sum_{x>\tau}\tilde{h}(x)f^{+}(x) = 0 \mbox{ and } \sum_{x>\tau}\tilde{h}^{2}(x)f^{+}(x) < \infty.
$$

In particular, note that such an $\tilde{h}$ is orthogonal to elements of $\mathcal{\dot{P}}_{F}^{+}$ itself. These, once again, have the form $\frac{(1-q)FG_0}{f}$ for some $G_0 \in \mathcal{G}$. By design, let us choose $G_0$ such that $G_{0}(x_1)=F(x_2)$ and $G_{0}(x_2)=-F(x_1)$ for $x_1, x_2$ in the support of $F$ and $G_{0}(x)=0$ elsewhere. It is easy to verify that such a choice does indeed lie within $\mathcal{G}$. On the other hand, the orthogonality of $\tilde{h}$ and  $\frac{(1-q)FG_0}{f}$ in $\LL^{2}(f^{+})$ implies that:
$$
\sum_{x\geq1}\frac{F(x)G_{0}(x)}{f(x)}\tilde{h}(x)f^{+}(x) = 0,
$$
or equivalently
$$
\sum_{x\geq1}F(x)G_{0}(x)\tilde{h}(x) = F(x_1)F(x_2)\left(\tilde{h}(x_1)-\tilde{h}(x_2)\right)=0.
$$


As $F(x)$ is strictly positive over its support, this implies that $\tilde{h}(x_1) - \tilde{h}(x_2) = 0.$ Thus all such $\tilde{h}$ must be constant on the support of $F$.


Now let us specialize $\tilde h$ to the components of the efficient score function, by writing them as $\tilde\ell = \dot\ell - \varPi_F\dot\ell$. Since we have thus determined that $\varPi_F\dot\ell$ vanishes on $x\leq\tau$ and $\tilde\ell$ is constant over $x>\tau$, we have therefore established the expression of the projection as in Equation \eqref{eq:proj} as claimed. To obtain the expression of the constant in Equation \eqref{eq:constant}, we can once again use the fact that $\varPi_F\dot\ell$ is a linear combination of $(1-q)FG_0/f$ for $G_0\in\mathcal{G}$, in addition to the fact that $\sum_x FG_0 = 0$ for all such $G$, to write:
$$
\sum_{x>\tau} (\dot\ell(x) - c) f(x) = \sum_x \varPi_F\dot\ell = \sum_x (1-q)F \int G = 0.
$$


Now, we can easily compute the efficient score functions using:
\begin{equation} \label{eq:scoreEff}
 \tilde{\ell}(x) = \begin{cases}
                      \dot{\ell}(x) \text{ if } 1\leq x \leq \tau, \\
                       c(\dot{\ell}) \text{ if } x > \tau.
                     \end{cases}
\end{equation}
using the expressions of $\dot{\ell}_{q}$ and $\dot{\ell}_{\theta}$ in Equation \eqref{Eq:scores}, we explicitly get $\tilde{\ell}_{q}$ and $\tilde{\ell}_{\theta}$. We start with $\tilde{\ell}_{q}$. We have:
\begin{eqnarray*}
 \sum_{x>\tau} \dot{\ell}_{q}(x)f(x) &=& \sum_{x>\tau}R_{\theta}(x) -1 +
 \frac{R_{\theta}(0)}{1-qR_{\theta}(0)}\sum_{x>\tau}f(x) \\
 &=& \frac{R_{\theta}(0)}{1-qR_{\theta}(0)}\sum_{x>\tau}f(x) - \sum_{x=0}^{\tau}R_{\theta}(x).
\end{eqnarray*}
We then determine $c(\dot{\ell}_{q})$ from Equation \eqref{eq:constant},
 \begin{eqnarray*}
  c(\dot{\ell}_{q}) &=& \frac{R_{\theta}(0)}{1-qR_{\theta}(0)} -
  \frac{\sum_{x=0}^{\tau}R_{\theta}(x)}{\sum_{x>\tau}f(x)}\\
  &=&\frac{R_{\theta}(0)}{1-qR_{\theta}(0)} -
  \frac{\sum_{x=0}^{\tau}R_{\theta}(x)}{1-q\sum_{x=0}^{\tau}R_{\theta}(x)}
 \end{eqnarray*}
and we finally obtain $\tilde{\ell}_{q}(x)$ as
  \begin{equation*}
   \tilde{\ell}_{q}(x) = \frac{1}{q} \ind\{x\leq \tau\} -
   \frac{\sum_{x=0}^{\tau}R_{\theta}(x)}{1-q\sum_{x=0}^{\tau}R_{\theta}(x)} \ind\{x> \tau\}+
   \frac{R_{\theta}(0)}{1-qR_{\theta}(0)}.
  \end{equation*}
Moving on to $\dot{\ell}_{\theta}$, from Equation \eqref{Eq:scores} and using the fact that $\sum_x \dot R_\theta(x)=0$, we have:
\begin{eqnarray*}
 \sum_{x>\tau} \dot{\ell}_{\theta}(x)f(x) &=& q\sum_{x>\tau}\dot{R}_{\theta}(x)  +
 \frac{\dot{R}_{\theta}(0)}{1-qR_{\theta}(0)}\sum_{x>\tau}f(x) \\
 &=& \frac{\dot{R}_{\theta}(0)}{1-qR_{\theta}(0)}\sum_{x>\tau}f(x) -
 q\sum_{x=0}^{\tau}\dot{R}_{\theta}(x).
\end{eqnarray*}
Then
\begin{equation*}
 c(\dot{\ell}_{\theta}) =\frac{q\dot{R}_{\theta}(0)}{1-qR_{\theta}(0)} -
  \frac{q\sum_{x=0}^{\tau}\dot{R}_{\theta}(x)}{1-q\sum_{x=0}^{\tau}R_{\theta}(x)}
\end{equation*}
 and
 \begin{equation*}
    \tilde{\ell}_{\theta}(x) = \frac{\dot{R}_{\theta}(x) }{R_{\theta}(x)} \ind\{x\leq \tau\}-
   \frac{q\sum_{x=0}^{\tau}\dot{R}_{\theta}(x)}{1-q\sum_{x=0}^{\tau}R_{\theta}(x)} \ind\{x> \tau\}+
   \frac{q\dot{R}_{\theta}(0)}{1-qR_{\theta}(0)}.
 \end{equation*}
 The  efficient Fisher information matrix has coefficients defined as
 \begin{equation} \label{defI}
 \tilde{I}_{q} = \sum_{x\geq1} \tilde{\ell}_{q}^{2}(x)f^{+}(x) \text{, }
 \tilde{I}_{q\theta_j} = \sum_{x\geq1}\tilde{\ell}_{q}(x) \tilde{\ell}_{\theta_j}(x)f^{+}(x)\text{ and }
\tilde{I}_{\theta_i \theta_j} = \sum_{x\geq1} \tilde{\ell}_{\theta_i}(x) \tilde{\ell}_{\theta_j}(x)f^{+}(x)
\end{equation}
for all $i,j = 1, \dots, k.$ Recall that when we write $\tilde\ell_\theta$, we are referring to a vector of score functions, whereas $\tilde{\ell}_{\theta_j}$ stands for the $j^{th}$ coordinate of $\tilde{\ell}_{\theta}$. The computation of these coefficients leads to
 \begin{equation} \label{Iqtilde}
  \tilde{I}_{q} = \frac{1}{1-qR_{\theta}(0)} \left\{ \frac{1}{q}\sum_{x=1}^{\tau} R_{\theta}(x) +
  \frac{[\sum_{x=0}^{\tau} R_{\theta}(x)]^{2}}{1-q\sum_{x=0}^{\tau} R_{\theta}(x)} -
  \frac{[R_{\theta}(0)]^{2}}{1-qR_{\theta}(0)}\right\}
 \end{equation}

  \begin{equation}\label{Ithetatilde}
  \tilde{I}_{\theta_i \theta_j} = \frac{q}{1-qR_{\theta}(0)} \left\{\sum_{x=1}^{\tau}
  \frac{[\dot{R}_{\theta}^{i}(x)][\dot{R}_{\theta}^{j}(x)]}{R_{\theta}(x)} +
  \frac{q[\sum_{x=0}^{\tau} \dot{R}_{\theta}^{i}(x)][\sum_{x=0}^{\tau} \dot{R}_{\theta}^{j}(x)]}
  {1-q\sum_{x=0}^{\tau} R_{\theta}(x)} -
  \frac{q[\dot{R}_{\theta}^{i}(0)][\dot{R}_{\theta}^{j}(0)] }{1-qR_{\theta}(0)}\right\}
 \end{equation}

 \begin{equation}\label{Iqthetatilde}
  \tilde{I}_{q\theta_j} = \frac{q}{1-qR_{\theta}(0)} \left\{ \frac{1}{q}\sum_{x=1}^{\tau} \dot{R}_{\theta}^{j}(x) +
  \frac{[\sum_{x=0}^{\tau} R_{\theta}(x)][\sum_{x=0}^{\tau}\dot{R}_{\theta}^{j}(x)]}{1-q\sum_{x=0}^{\tau} R_{\theta}(x)} -
  \frac{\dot{R}_{\theta}^{j}(0)R_{\theta}(0)}{1-qR_{\theta}(0)}\right\}
 \end{equation}
 with $\dot{R}_{\theta}^{j}$ the partial derivative of $R_{\theta}$ with respect to the $j^{th}$ coordinate of $\theta.$
\cqfd

\bigskip

\subsection{Proof of Theorem~\ref{TheoEfficiency}\\}

We first state and prove two lemmas that will be used for the proof of Theorem~\ref{TheoEfficiency}. \\

As usual, let $\alpha$ be a component of the parameters vector $(q,\theta)$, denote by $\alpha_0$ the true value of $\alpha$ (if it exists), and let $v(\alpha_0)$ be a closed neighborhood of $\alpha_0$. We denote by $\h_{\alpha}$ the subset of $\LL^{2}(f^{+})$ defined by
 \begin{equation}\label{eq5}
  \h_{\alpha} = \left\{ \tilde{\ell}_{\alpha}, \text{ with } \alpha \in v(\alpha_0)\right\}.
 \end{equation}

\begin{lem} \label{Donsker}
Let $\hat{\alpha}$ be a consistent estimator of $\alpha_0.$ If $\theta \mapsto R_\theta(x)$ is twice continuously differentiable for every $x\leq\tau$ and Assumptions $1$-$3$ hold, then $\h_{\alpha}$ is a Donsker class with square integrable envelope that contains $\tilde{\ell}_{\hat{\alpha}}$ with probability that tends to one.
\end{lem}
\paragraph{Proof \\}

We adapt the method used in Example $19.7$ from~\cite{AWV98}. Recall that a $\delta$-bracket is a subset $[u_1,u_2]$ of $\LL^{2}(f^{+})$ such that  $\| u_2-u_1\|_{\LL^{2}(f^{+})} < \delta.$ The bracketing number $ N\!\left(\delta,\h_{\alpha}, \LL^{2}(f^{+}) \right)$ is the minimum number of $\delta$-brackets needed to cover $ \h_{\alpha}$ and the bracketing entropy is the logarithm of this quantity. To show that $\h_{\alpha}$ is Donsker, we establish the sufficient condition that the square root entropy integral
\begin{equation} \label{bracket_I}
  \int_{0}^{1}\sqrt{\log N\left(\gamma,\h_{\alpha}, \LL^{2}(f^{+}) \right)}d\gamma
\end{equation}
is finite. (See Theorem $19.5$ in~\cite{AWV98}.)

We begin by establishing continuity properties of the parametric efficient score functions. From the differentiability of $R_\theta$ in $\theta$ and the expressions given in Proposition \ref{Ef_Scor}, it is evident that $\tilde\ell_\alpha$ is always a differentiable function of $\alpha$. Let us denote these derivatives by $\dot{\tilde{\ell}}_\alpha$. For $\alpha=q$ and $\alpha=\theta_j$ we can respectively compute these as
$$
 \dot{\tilde{\ell}}_q=\left\{\left( \frac{R_{\theta}(0)}{1-qR_{\theta}(0)}\right)^{2} -\frac{1}{q^{2}}\right\}\ind\{x\leq \tau\} +
 \left\{\left( \frac{R_{\theta}(0)}{1-qR_{\theta}(0)}\right)^{2} -
 \left( \frac{\sum_{k=0}^{\tau} R_{\theta}(k)}{1-q\sum_{k=0}^{\tau} R_{\theta}(k)} \right)^{2} \right\}\ind\{x> \tau\}
$$
and
$$
  \dot{\tilde{\ell}}_{\theta_j}=\left\{ \frac{\ddot{R}_{\theta}^{j}(x)}{R_{\theta}(x)} -
  \left(\frac{\dot{R}_{\theta}^{j}(x)}{R_{\theta}(x)}\right)^{2} + \frac{q\ddot{R}_{\theta}^{j}(0)}{1-qR_{\theta}(0)}
  + \left(\frac{q\dot{R}_{\theta}^{j}(0)}{1-qR_{\theta}(0)}\right)^{2}\right\}\ind\{x\leq \tau\} + $$$$
  \left\{ \frac{q \ddot{R}_{\theta}^{j}(0)}{1-qR_{\theta}(0)} + \left(\frac{q\dot{R}_{\theta}^{j}(0)}{1-qR_{\theta}(0)}\right)^{2}
  -\frac{q\sum_{k=0}^{\tau}\ddot{R}_{\theta}^{j}(k)}{1-q\sum_{k=0}^{\tau}R_{\theta}(k)} -
  \left(\frac{q\sum_{k=0}^{\tau}\dot{R}_{\theta}^{j}(k)}{1-q\sum_{k=0}^{\tau}R_{\theta}(k)}\right)^{2} \right\}\ind\{x> \tau\}.
$$

By inspection, we find that $\dot{\tilde{\ell}}_q$ is always continuous itself, and that $\dot{\tilde{\ell}}_{\theta_j}$ is also continuous provided that $\theta\mapsto\R_\theta$ is in $\mathcal{C}^{2}$ and $R_\theta(x)\geq\eta>0$ for all $x\leq\tau$, as assumed. These conditions also imply that $\dot{\tilde{\ell}}_\alpha$ have a finite $\LL^{2}(f^{+})$-norm and that these functions are Lipschitz-continuous on $v(\alpha_0)$. We thus have a non-negative bounded $V$ such that
 \begin{equation} \label{lipCond}
  | \tilde{\ell}_{\alpha_2}(x) - \tilde{\ell}_{\alpha_1}(x)| \leq V|\alpha_2 - \alpha_1|
  \text{ for every } \alpha_1, \alpha_2 \in v(\alpha_0).
 \end{equation}

Now, from this Lipschitz condition, it follows that if $|\alpha-\alpha_1| < \epsilon$ then $\tilde{\ell}_{\alpha_1} - \epsilon V \leq \tilde{\ell}_{\alpha} \leq   \tilde{\ell}_{\alpha_1} +\epsilon V.$ This means that we need as many $\epsilon$-balls (a ball with radius $\epsilon/2$) to cover $v(\alpha_0)$ as we need $\delta$-brackets $(\delta = 2\epsilon V)$ to cover $\h_{\alpha}.$ Since the number $n_0$ of $\epsilon$-balls needed to cover  $v(\alpha_0)$ is such that
$$
  n_0 \leq C \frac{\mathsf{diam}[v(\alpha_0)]}{\epsilon} \vee 1,
$$
with $C$ a constant depending only on $v(\alpha_0)$, it follows that the bracketing number is
\begin{equation} \label{bracket_N}
  N\!\left(\delta,\h_{\alpha}, \LL^{2}(f^{+}) \right) \leq \left\{ C \mathsf{diam}[v(\alpha_0)]
  \frac{2V}{\delta}\right\} \vee 1.
\end{equation}
Thus the bracketing entropy is of order smaller than $\log(1/\delta)$, whose square root is integrable near $0$. This establishes the sufficient condition of Equation \eqref{bracket_I}, and thus $\h_\alpha$ is indeed Donsker.


To complete the other claims of the proof, note that for all $\alpha$ in $v(\alpha_0),$ $\tilde{\ell}_{\alpha}$ has a finite $\LL^{2}(f^{+})$-norm and that $|\tilde{\ell}_{\alpha} (x)| \leq U$ for some $U<\infty$, for all $x\geq 1$. The boundedness of $\tilde{\ell}_{\alpha}$ is obtained from the expression of $\tilde{\ell}_{q}$ and $\tilde{\ell}_{\theta_j}.$ The constant function $U$ is a square integrable envelope for  $\h_{\alpha}.$  We use  the continuity of the map $\alpha \mapsto \tilde{\ell}_{\alpha}(x)$ and consistency of $\hat{\alpha}$ to show that  $ \lim_{N\rightarrow \infty} \po [| \tilde{\ell}_{\hat{\alpha}}(x)- \tilde{\ell}_{\alpha_0}(x)|> \epsilon] = 0$ for all $x\geq 1$. This proves that $\h_{\alpha}$ contains  $\tilde{\ell}_{\hat{\alpha}}$ with probability that tends to one and the lemma holds. \cqfd
 
\bigskip
  
The result in Lemma~\ref{Donsker} holds for $ \h_{q}$ and $\h_{\theta_j}$ for all $j=1,\dots,k$ and thus also for their union $\h.$ We conclude that $\h$ is a Donsker class with square integrable envelope that contains $(\tilde{\ell}_{\hat{q}},\tilde{\ell}_{\hat{\theta}})$ with probability that tends to one.

\begin{lem}\label{lemma:score-equations}
  $\hat{\theta}_\tau$ and $\hat{q}_\tau$ solve the efficient score equations:
  \begin{equation} \label{Def_CSPE12}
    \sum_{i=1}^{D}\tilde{\ell}_{q}(x_i) = 0, \mbox{ and } \sum_{i=1}^{D}\tilde{\ell}_{\theta}(x_i) = 0.
  \end{equation}
\end{lem}
\paragraph{Proof}
Note that the efficient score equation $ \sum_{i=1}^{D}\tilde{\ell}_{q}(x_i) = 0$ leads to equality
\begin{equation} \label{Eq:mleq}
	\frac{D_{\tau}}{q} + \frac{DR_{\theta}(0)}{1-qR_{\theta}(0)}
	-\frac{(D-D_{\tau})\sum_{k=0}^{\tau}R_{\theta}(k)}{1-q\sum_{k=0}^{\tau}R_{\theta}(k)} = 0
\end{equation}
whose solution is  $\hat{q}(\theta)$ given by
\begin{equation} \label{mleq}
	\hat{q}(\theta) = \frac{D_{\tau}}{D\sum_{k=1}^{\tau}R_{\theta}(k) + D_{\tau}R_{\theta}(0) }.
\end{equation}
 Likewise, if one sets to zero all the partial derivatives
of the logarithm of the likelihood in \eqref{TruncVrais}, one has
\begin{equation} \label{EqTheta}
  \sum_{x=1}^{\tau}\frac{\dot{R}_{\theta}(x)}{R_{\theta}(x)}n_{x} -
D_{\tau} \frac{\sum_{k=1}^{\tau}\dot{R}_{\theta}(k)}{\sum_{k=1}^{\tau}R_{\theta}(k)} = 0.
\end{equation}
This equality is equivalent to $\sum_{i=1}^{D}\tilde{\ell}_{\theta}(x_i) = 0$ with $q$ replaced by
$\hat{q}(\theta)$ in $\tilde{\ell}_{\theta}.$ The zero notation here refers to the null vector of $\R^k$. \cqfd

We now prove the asymptotic efficiency of the estimators. Note that all results in this proof are stated under the restriction $\tau \leq \tau_{0}$ when necessary.

By Lemma \ref{lemma:score-equations}, $\hat{\theta} $ and $\hat{q}(\hat{\theta})$ are such that
\begin{equation} \label{eq1}
\frac{1}{\sqrt{D}}\sum_{i=1}^{D}\tilde{\ell}_{\hat{\theta}}(x_i) = 0 \text{ and }
\frac{1}{\sqrt{D}}\sum_{i=1}^{D}\tilde{\ell}_{\hat{q}}(x_i) = 0.
\end{equation}
 
As $\tilde{\ell}_{\theta}$ and $\tilde{\ell}_{q}$ are free of $F$ and that this is also true for the plug-in estimators  $\tilde{\ell}_{\hat{\theta}}$ and $\tilde{\ell}_{\hat{q}}$, it is not difficult to verify  that
\begin{equation} \label{eq2}
  \sum_{x\geq1}\tilde{\ell}_{\hat{\theta}}(x)f_{(\hat{q},\hat{\theta},F)}^{+}(x) = 0 \text{ and }
  \sum_{x\geq1}\tilde{\ell}_{\hat{q}}(x)f_{(\hat{q},\hat{\theta},F)}^{+}(x) = 0.
\end{equation}
The asymptotic efficiency of $(\hat{q},\hat{\theta})$ follows from Theorem $25.54$ in~\cite{AWV98}. As assumptions, this theorem needs the assertions of Theorem~\ref{Consistency} (consistency) and Lemma~\ref{Donsker} (Donsker property), in addition to the following two convergence properties pertaining to the ``plug-in'' score functions. In particular, we need to show that our estimators $(\hat{q},\hat{\theta}):$ satisfy:
\begin{equation}\label{eq3}
  \| \tilde{\ell}_{(\hat{q},\hat{\theta})} -
  \tilde{\ell}_{(q_0,\theta_0)}\|_{\LL^{2}(f^{+})}^{2} = \text{o}_\po(1),
\end{equation}
and
\begin{equation}\label{eq4}
  \| \tilde{\ell}_{(\hat{q},\hat{\theta})}\|_{\LL^{2}(\hat{f}^{+})}^{2} = \sum_{x\geq1}    \| \tilde{\ell}_{(\hat{q},\hat{\theta})}(x) \|_{2}^{2} \hat{f}^{+}(x) = \text{O}_\po(1),
\end{equation}
where $f^{+}$ stands for $f_{(q_0,\theta_0,F)}^{+}$, $\hat{f}^{+}$ stands for the parametric plug-in $f_{(\hat{q},\hat{\theta},F)}^{+}$, and $\tilde{\ell}_{(q_0,\theta_0)}$ and $\tilde{\ell}_{(\hat{q},\hat{\theta})}$ are the stacked vectors of $(k+1)$ components, $(\tilde{\ell}_{q_0},\tilde{\ell}_{\theta_0})$ and $(\tilde{\ell}_{\hat{q}},\tilde{\ell}_{\hat{\theta}})$ respectively.

To establish Equations \eqref{eq3} and \eqref{eq4}, we can use for each parameter $\alpha$ the continuity properties of $\tilde{\ell}_\alpha$ per component, as in the proof of Lemma \ref{Donsker}. In particular, note first that for all $x>\tau_0$, we have that $\tilde \ell_\alpha(x)$ is constant. Therefore, for each parameter $\alpha$ we need only to account for the convergence of $\tilde\ell_{\hat{\alpha}}(x)\to\tilde\ell_\alpha(x)$ for $x=1,\cdots,\tau_0+1$, all of which happen (in probability), by continuity.

It follows that for each $x$, $\| \tilde{\ell}_{(\hat{q},\hat{\theta})}(x) - \tilde{\ell}_{(q_0,\theta_0)}(x)\|^2$ converges to $0$ in probability, and since we have only finitely many distinct values, the convergence is uniform for all $x$. Equation \eqref{eq3} is thus immediate.

On the other hand, $\hat{f}^+(x)\to f^+(x)$ in probability for each $x$. By finiteness, it follows that $\sum_{x\leq\tau_0} \hat{f}^+(x) \to \sum_{x\leq\tau_0} f^+(x)$, and consequently $\sum_{x>\tau_0} \hat{f}^+(x) \       to \sum_{x>\tau_0} f^+(x)$. By using once again the fact that $\tilde\ell$ is constant beyond $\tau$, the convergence reduces again to finitely many convergences, and thus $\sum_{x} \|\tilde\ell_{(q_0,\theta_0)}(x)\|_2^2 \hat{f}^+(x) \to \sum_{x} \|\tilde\ell_{(q_0,\theta_0)}(x)\|_2^2 f^+(x)$. We can therefore write:
\begin{eqnarray*}
\sum_{x}    \| \tilde{\ell}_{(\hat{q},\hat{\theta})}(x) \|_{2}^{2} \hat{f}^{+}(x)
&\leq& 2\sum_{x} \| \tilde{\ell}_{(\hat{q},\hat{\theta})}(x)-\tilde\ell_{(q_0,\theta_0)}(x) \|_{2}^{2} \hat{f}^{+}(x) + 2\sum_x \|\tilde\ell_{(q_0,\theta_0)}(x)\|_2^2 \hat{f}^{+}(x)\\
&=& \text{o}_\po(1)+\text{O}_\po\left(\|\tilde{\ell}_{(q_0,\theta_0)}\|_{\LL^{2}(f^{+})}^{2} \right),
\end{eqnarray*}
which completes the proof of Equation \eqref{eq4} and the theorem. \cqfd

\bibliographystyle{authordate1}
\bibliography{Biblio}
\end{document}